\documentclass[aps,twocolumn,prd,superscriptaddress,showpacs,preprintnumbers,nofootinbib]{revtex4-1}

\usepackage{mathrsfs}

\usepackage{dsfont}
\usepackage{graphicx}
\usepackage{slashed}
\usepackage{amsmath,amssymb,amsthm,amsfonts}
\usepackage{dsfont,bbm}
\usepackage{color,rotating}
\usepackage{hyperref}
\usepackage[caption=false]{subfig}
 \def\Eq#1{Eq.~(\ref{#1})}
\def\Fig#1{Fig.~\ref{#1}} 

\def\0#1#2{\frac{#1}{#2}}
\def\eq#1{(\ref{#1})}

\newcommand{\imag}{\text{i}}
\newcommand{\skipthis}[1]{}
\renewcommand{\d}{{\text{d}}}
\newcommand{\Tr}{{\text{Tr}}}
\newcommand{\sign}{{\text{sign}}}

 \def\Eq#1{Eq.~(\ref{#1})}
\def\Fig#1{Fig.~\ref{#1}}

\def\Fig#1{Fig.~\ref{#1}}

\def\s0#1#2{\mbox{\small{$ \frac{#1}{#2} $}}}
\def\0#1#2{\frac{#1}{#2}}

\newcommand{\sumint}{\int\hspace{-4.8mm}\sum}

\newcommand{\beq}{\begin{equation}}
\newcommand{\eeq}{\end{equation}}
\newcommand{\beqa}{\begin{eqnarray}}
\newcommand{\eeqa}{\end{eqnarray}}

\graphicspath{{./figures/}}

\newcommand{\bea}{\begin{eqnarray}}
\newcommand{\eea}{\end{eqnarray}}

\definecolor{darkgreen}{rgb}{0,0.6,0}
\definecolor{gray}{rgb}{.7,.7,.7}

\def\eq#1{(\ref{#1})}
\def\Eq#1{Eq.~(\ref{#1})}
\newcommand {\apgt} {\ {\raise-.5ex\hbox{$\buildrel>\over\sim$}}\ }
\newcommand {\aplt} {\ {\raise-.5ex\hbox{$\buildrel<\over\sim$}}\ }

\def\s0#1#2{\mbox{\small{$ \frac{#1}{#2} $}}}
\def\0#1#2{\frac{#1}{#2}}

\def\N{\mathds{N}}

\def\CC{{\mathcal C}}

\def\CL{{\mathcal L}}

\def\CT{{\mathcal T}}

\newcommand{\be}{\begin{eqnarray}}
\newcommand{\ee}{\end{eqnarray}}

\begin{document}

\title{Real time correlation functions and the functional renormalisation group}

 \author{Jan M. Pawlowski}
  \affiliation{Institut f\"ur Theoretische
  Physik, Universit\"at Heidelberg, Philosophenweg 16, 69120
  Heidelberg, Germany} 
\affiliation{ExtreMe Matter Institute EMMI, GSI, Planckstr. 1,
  D-64291 Darmstadt, Germany}

\author{Nils Strodthoff}
\affiliation{Institut f\"ur Theoretische
  Physik, Universit\"at Heidelberg, Philosophenweg 16, 69120
  Heidelberg, Germany}

\pacs{12.38.Aw, 
11.10.Wx	, 
11.30.Rd	, 
12.38.Gc}		
\begin{abstract}
  We put forward a functional renormalisation group approach for the
  direct computation of real time correlation functions, also
  applicable at finite temperature and density. We construct a general
  class of regulators that preserve the space-time symmetries, and
  allows the computation of correlation functions at complex
  frequencies. This includes both imaginary time and real time, and
  allows in particular the use of the plethora of imaginary time
  results for the computation of real time correlation functions.  We
  also discuss real time computation on the Keldysh contour with
  general spatial momentum regulators. Both set-ups give access to the
  general momentum and frequency dependence of correlation functions.
\end{abstract}
\maketitle
\section{Introduction}
Real time correlation functions and most importantly real time
propagators are the key ingredient to gain access to dynamical
observables in strongly interacting quantum field theories. One
prominent example is the single particle spectral function that is
defined via the imaginary part of the retarded two-point function. It
comprises information about the spectrum of the theory and in
particular about resonances, and their fate at finite temperature and
density. Single particle spectral functions also serve as an input for
the calculation of transport coefficients via Kubo formulae in an
approach put forward recently in
\cite{Haas:2013hpa,Christiansen:2014ypa}.

In strongly correlated regimes of the theory at hand already imaginary
time computations are quite involved. A prominent example is QCD,
where first principle approaches, both in the continuum and on the
lattice, only provide numerical access to the low-energy hadronic
regime with confinement and chiral symmetry breaking. 
For example, in continuum QCD
elaborated approximations, including the implementation of symmetry
constraints, and solving techniques have been developed for the
qualitative and quantitative access to QCD correlation
functions \cite{Mitter:2014wpa,Braun:2014ata}. 
Moreover, in most cases correlation functions can only be
computed numerically. 
Then, an analytic continuation from Euclidean (imaginary) space-time
to Minkowski space-time has to be carried out based on numerical Euclidean
data. This reconstruction of real time correlation functions such as
spectral functions using given numerical Euclidean data is usually
done with Maximum Entropy methods, Pad\'{e} approximants or similar
reconstruction methods. However, these approaches imply a certain bias
about the continuation, require very accurate Euclidean data, and hence 
represent a challenging conceptual and numerical problem.

The above reconstruction problem can only be overcome within a direct
numerical real time computation. The most direct approach is a
computation within a real time formulation of the theory. This entails
that the nontrivial strongly correlated dynamics of the theory has to
be accessed within such a real time approach. Even though possible in
principle, this does not allow to utilise directly the plethora of
results obtained in Euclidean space for example in the form of numerical input data. An alternative approach that naturally
extends and utilises the Euclidean computation to real time
correlation functions is the extension of the former to complex
frequencies. In a functional approach this is, loosely speaking, based
on the computation of Euclidean momentum loops at complex external
frequencies. This includes both, imaginary frequencies or Euclidean
space-time and real frequencies or Minkowski space-time. Such an
approach was put forward in the functional 
renormalisation group (FRG),
\cite{Floerchinger:2011sc,Strodthoff:2011tz,Kamikado:2012bt,
  Kamikado:2013sia,Tripolt:2013jra,Tripolt:2014wra}, for recent work
with Dyson Schwinger equations (DSE) see \cite{Windisch:2012zd,Strauss:2012dg,%
Windisch:2012sz,Windisch:2013mg,Windisch:2013dxa}. 

Here we put forward a general FRG-framework applicable in the presence
of general regulators and for full frequency and momentum dependencies
of the correlation functions. This is indispensable in situations with
nontrivial quantum, thermal and medium corrections to the dispersion
relations of the theory at hand. The latter includes a wide range of applications  
from condensed matter, heavy ion collisions to quantum gravity. 

More specifically we aim at calculations of the hadron spectrum and
real time observables in general within the framework of the fQCD
collaboration \cite{fQCD}. The fQCD collaboration works on a
quantitative first-principle approach to continuum QCD within the FRG
framework, see \cite{Mitter:2014wpa,Braun:2014ata} for first
publications. On the one hand, the application to quark and gluon
spectral functions requires an extension which goes beyond the widely
used spatial regulator functions. On the other hand, we aim at a fully
numerical procedure which is not tailored to specific regulator
classes. It has to work for generalisations of standard regulators
with exponential decay properties in order to make the problem
numerically tractable.

A further long-term goal is the extension of imaginary time results,
such as those obtained in the fQCD framework, to nonequilibrium
situations. There, the dynamics singles out the time direction in the
first place. This asks for approximation schemes and regulators that
take into account the causality constraints and conservation laws at
nonequilibrium. Related approximation schemes on the basis of the 2PI
approach have e.g.\ been discussed in
\cite{Blaizot:2010zx,Carrington:2014lba,Rentrop:2015tia}, causal (time) regulators
have been introduced in \cite{Gasenzer:2008zz} on the Keldysh
contour. In higher dimensions the latter can be amended by a regulator
in spatial momenta in a mixed representation. In the present work we
discuss the properties of such generic spatial momentum regularisation
and results for the single particle spectral functions.

The set-up for our approach is discussed in detail in the first
section of the paper and demonstrated using the spectral functions in
the $O(N)$ model as an illustrative example. The complementary second
part addresses different approaches towards direct real time
calculations, where the complications of Euclidean or Semi-Euclidean
approaches as the ones from above due to the necessity of performing
an analytic continuation are absent. Here we put forward the
formalism for the calculation of spectral functions in a closed time
path (CTP) framework.
\section{Propagators and single particle spectral functions}
Real- and imaginary time propagators are limiting cases of the
two-point correlation function or propagator with complex frequency
$\omega\in \mathbb{C}$.  The current framework is based on the
Euclidean imaginary time quantum field theory and we denote $\omega=
\omega_{\text{\tiny{E}}}+\imag \omega_{\text{\tiny{M}}}$ with
Euclidean frequency $\omega_{\text{\tiny{E}}}$ and Minkowski frequency
$\omega_{\text{\tiny{M}}}$. In this section we discuss the numerical
computation of correlation functions at complex frequencies from
Euclidean loop integrals. In the present work, we concentrate on the
important example of the single-particle spectral function, but the
formulation also applies to higher correlation functions.

\subsection{Spectral functions from Euclidean correlation functions at
  complex frequencies}
\label{sec:poleprocedures}
In the following we put forward an approach for the analytical
continuation of Euclidean correlation functions to complex
frequencies. The continuation is chosen such that they reduce to the
corresponding real time correlation functions at purely imaginary
frequency. The appropriate continuation is discussed at the example of
the propagator.

We start by defining a propagator $G(\omega, \vec p)$ for complex
frequencies $\omega=\omega_{\text{\tiny{E}}}+\imag
\omega_{\text{\tiny{M}}}$ as the uniquely defined analytic
continuation of the real time Feynman propagator $G_{\text{F}}$ to
complex frequencies $\omega_{\text{\tiny{M}}}\in\mathbbm{C}$, i.e.\
\begin{align}
\label{eq:defG}
G(\imag \omega_{\text{\tiny{M}}},\vec p):&=-G_{\text{F}}(
\omega_{\text{\tiny{M}}},\vec p)\nonumber\\[2ex]
&=\imag \int \d^4 x \,\langle \CT \phi(x)\phi(0)\rangle_c\,
e^{-\imag\omega_{\text{\tiny{M}}} x^0+\imag\vec x \vec p}\,,
\end{align}
where $\CT$ denotes time ordering and the subscript indicates the
connected two-point correlation function. \Eq{eq:defG} also holds true
at finite temperature and density. Then the real time retarded
propagator can be obtained from
\begin{equation}
\label{eq:EuclideanRetardedGamma2}
G_{\text{R}}(p_0,\vec p)=-\lim_{\epsilon\to 0} G(-\imag(p_0+\imag \epsilon),\vec p)\,,
\end{equation}
that is from the propagator evaluated at the complex momentum
$\epsilon - \imag p_0$ with $\omega_{\text{\tiny{E}}}=\epsilon$ and
$\omega_{\text{\tiny{M}}}=-p_0$. The imaginary part of the retarded
propagator relates directly to the spectral function,
\begin{align}
\label{eq:spectraldef}
\rho(p)&=-2\,\text{Im}\, G_{\text{R}}(p)\,,
\end{align}
see \eq{eq:spectraldefapp}. Evaluated at Euclidean frequencies
$\omega=2\pi n T$ with $n\in\mathbbm{Z}$ the continuation $G$ defined
in \eq{eq:defG} coincides with the Euclidean Feynman propagator,
$G(\omega,\vec p)=G_\text{E}(\omega,\vec p)$, which is defined via
\begin{align}\label{eq:propcomplex} 
  G_\text{E}(\omega,\vec p):=\int_0^{\beta}\hspace{-0.2cm}  \d x^0 \int \d^3 x\, \langle
  \CT \phi(x)\phi(0)\rangle^{\ }_c\, e^{\imag (\omega x_0
    +\vec p \vec x)}\,.
\end{align} 
Note that at finite temperature the Euclidean propagator
\eq{eq:propcomplex} is only defined for discrete values $\omega=2\pi n
T$ with $n\in\mathbbm{Z}$. Therefore, there is no unique analytic
continuation to $\omega\in\mathbbm{C}$. A perfectly well-defined
continuation is defined by taking \eq{eq:propcomplex} for
$\omega\in\mathbbm{R}$ on the whole real axis which has a unique
continuation to $\omega\in\mathbbm{C}$. The continuation defined in
this way is a straightforward one in a Euclidean framework, and
can be applied to lattice correlation functions in position space. 

However, this continuation does not coincide with the real time
propagator at complex momenta. Moreover, in continuum frameworks
quantum corrections to correlation functions are computed from loop
representations in frequency and momentum space, e.g.\ in perturbation
theory and in particular in the method used in the present work, the
FRG. Then, the frequency routing is non-unique for $\omega\neq 2\pi n
T$. This ambiguity can be used to our advantage in order to define the
continuation to complex frequencies in such a way, that the difference
to real time correlation functions is easily tracked down. Indeed, for
any loop representation in frequency space the difference can be
written as a sum over residues of the poles that depend on the
external frequency, as well as potential cuts. For the Feynman
propagator this reads
\begin{equation}
\label{eq:ineqcontinuations}
G_{\text{F}}(\omega_{\text{\tiny{M}}})=-G_{\text{E,\tiny{cont}}}(\imag \omega_{\text{\tiny{M}}})
+\sum_{\rm poles} {\rm Res}+{\rm Cuts}\,, 
\end{equation}
see \eq{eq:DeltaGeuc} for an explicit one-loop example. Note that the
frequency routing can be chosen separately for each loop in order to
facilitate the computation. To sum up, the only necessary properties
for a given analytic continuation of Euclidean correlation functions
are (i) the equivalence on the Matsubara frequencies $\omega = 2\pi n
T$, and (ii) the difference to the real time correlation function is
accessible, see \eq{eq:ineqcontinuations}.  Hence, as (i) is trivially
satisfied, the relation \eq{eq:ineqcontinuations} is at the heart of a
numerical computation of the real time propagators, and in particular
the spectral function.

Thus, the key step is the computation of $G(- \imag \omega,\vec p)$ at complex
frequencies $\omega\in \mathbbm{C}$ from the Euclidean propagator
$G_{\text{E,\tiny{cont}}}(\omega,\vec p)$.  The latter is directly accessible in the
present FRG framework. The difference is entailed in
\eq{eq:ineqcontinuations}, and the pole positions and the corresponding 
residues can also be computed numerically. 

The numerical FRG-computation of the propagator $G_{\text{E,\tiny{cont}}}(\omega,\vec p)$ is 
facilitated by the fact, that all FRG-relations for correlation functions are one-loop
exact. This is very amiable for computations at complex external
frequency, as one only has to numerically follow poles and cuts at one
loop. It also allows to discuss the important properties of such a
procedure already at the perturbative one-loop example with classical
propagators and vertices. The additional pole and cut structures
arising from the introduction of momentum- and frequency-dependent
regulators are then discussed in a second step.

Hence, we first consider the one-loop correction to the Euclidean
propagator $\Delta \Gamma^{(2)}$ that arises from a loop diagram with
three-point vertices $\Gamma^{(3)}$. The latter are assumed to be
momentum-independent, and we allow for two different propagators $G_1$
and $G_2$, corresponding to masses $m_1$ and $m_2$ respectively.  This
already encompasses the important case of loops with radial modes
$\sigma$ and (pseudo-) Goldstone modes $\vec \pi$. In low-energy
effective theories of QCD these stand for the lowest
scalar-pseudo-scalar meson multiplet, the sigma meson and the
pions. The corresponding one-loop contribution for Euclidean external
frequencies reads
\begin{align}\nonumber 
  \Delta \Gamma^{(2)}_{\text{E,\tiny{cont}}}(p)=&[\Gamma^{(3)}]^2 \sumint_q G_1(q)
  G_2(p+q)\\[2ex]
  =&[\Gamma^{(3)}]^2\sumint_q\frac{1}{q_0^2+(\epsilon^1_{q})^2}
\frac{1}{(q_0+p_0)^2
    +(\epsilon^2_{q+p})^2}\,,
\label{eq:preDeltaGeuc}\end{align}
with constant classical vertex $\Gamma^{(3)}$ and the quasiparticle energies  
\begin{align}\label{eq:epsilon}
 \epsilon^i_q=\sqrt{\vec q^2+m_i^2}\,. 
\end{align}  
In \eq{eq:preDeltaGeuc} and in the following we use the shorthand
notations
\begin{align}
  \int_q\equiv\int \0{\d^4 q}{(2\pi)^4}\,,\quad \sumint_q\equiv
  T\sum_{q_0}\int \0{\d^3 q}{(2\pi)^3}\,,\quad \int_{\vec q}\equiv\int
  \0{\d^3 q}{(2\pi)^3}\,.
\end{align}
In \eq{eq:preDeltaGeuc} we have chosen the frequency routing in the
diagram such that one of the propagators, $G_1$, only depends on the loop
frequency $q_0  = 2 \pi n T$ with $n\in\mathbb{Z}$. For external
Matsubara frequencies $p_0= 2 \pi n T$ this agrees with the
routing $G_1(q-p)\,G_2(q)$ and other routings that are obtained by
shifts with Matsubara frequencies. For $p_0\neq 2 \pi n T$ all these
choices are different. Moreover, the analytic continuation specified
with the frequency routing in \eq{eq:preDeltaGeuc} does not coincide
with \eq{eq:propcomplex} for $p_0\neq 2 \pi n T $ with
$n\in\mathbbm{Z}$. This also holds for any other frequency routing. 

In the present perturbative one-loop example the above frequency
routing is very ad hoc as there is not even a selection criterion for
choosing $G_1(q)$ instead of $G_2(q)$. In the functional
renormalisation group approach diagrams such as \eq{eq:preDeltaGeuc}
are hit by derivatives w.r.t.\ the infrared cutoff scale that only act
on the propagators, see \eq{eq:dRk}. For diagrams with $n$ propagators
this leads to $n$ diagrams instead of one, and singles out exactly one
propagator in each of these diagrams. This provides the selection
criterion which makes the above frequency routing uniquely
defined. Importantly, this specific routing also simplifies the
numerics.

Evidently \eq{eq:preDeltaGeuc} satisfies the first of the two
properties of an analytic continuation defined below
\eq{eq:ineqcontinuations}, it agrees with the Euclidean result for
external Matsubara frequencies. As argued above, it is also uniquely
defined, at least in the functional renormalisation group approach.
This is an important prerequisite for the second property, access to
the difference between the real time correlation functions and the
specific analytic continuation of the Euclidean ones. This requires a
discussion of the pole structure of \eq{eq:preDeltaGeuc}. With the
bosonic thermal distribution
\begin{align}\label{eq:nb}
  n(\omega_{\text{\tiny{M}}}) =\0{1}{e^{\beta \omega_{\text{\tiny{M}}}}-1}\,,
\end{align}  
for Minkowski frequencies we can rewrite \eq{eq:preDeltaGeuc} in terms
of a contour integral surrounding the Euclidean axis. The Matsubara
frequencies are then the positions of the poles at
$\omega_{\text{\tiny}{E}}= 2 \pi n T$ with $n\in\mathbb{Z}$. For
Euclidean frequencies $p_0\in\mathbb{R}$ the integral is easily
performed and we arrive at
\begin{align}\nonumber 
  &\Delta \Gamma^{(2)}_{\text{E,\tiny{cont}}}(p)\\[2ex] 
\nonumber &=\frac{[\Gamma^{(3)}]^2}{2
    \imag}\sum_\pm\int_{\vec q} \Bigl(\underset{\pm\imag
    \epsilon_{q}^1}{\text{Res}}[G_1 G_2]
  \cdot\left[1+ 2 n(\mp\epsilon_{q}^1)\right]\\
  &+\underset{- p_0\pm\imag \epsilon_{q+p}^2}{\text{Res}}[G_1
  G_2]\cdot \left[1+ 2 n(-\imag
    p_0\mp\epsilon_{q+p}^2)\right]\Bigr)\,.
\label{eq:DeltaGeuc}\end{align}
The expression $\text{Res}_{r_0}[G_1 G_2]$ in \eq{eq:DeltaGeuc}
denotes the residue of the integrand $G_1(q) G_2(p+q)$ at
$q_0=r_0$. Coming back to the discussion which lead to \eq{eq:ineqcontinuations},
to obtain the contribution one requires the proper analytic
continuation of \eq{eq:DeltaGeuc} to imaginary momenta, which
is not uniquely defined at finite temperatures, where the propagator is
only given at discrete Matsubara frequencies.
Then, the additional requirement of appropriate
analyticity conditions \cite{Baym:1961} singles out a unique
continuation. This ambiguity is already visible in \eq{eq:DeltaGeuc},
where the omission of the external momentum in the argument of the
second distribution function leads to a different continuation which
coincides with \eq{eq:DeltaGeuc} for $p_0\in2\pi\mathbb{Z}$. The
desired continuation to obtain the retarded correlation function
can be determined within a real time formalism such as the closed time
path formalism discussed below.  Here the real time result is given by
\eq{eq:DeltaGeuc} omitting the external momentum in the argument of
the distribution function. This result coincides with that obtained
from the standard continuation procedure in the imaginary time
formalism \cite{Landsman:1986uw,LeBellac:2000}, where the periodicity
of the distribution functions for Euclidean external momenta is
exploited before the analytic continuation is performed.

\begin{figure}[t]
\centering
\includegraphics[width=0.7\columnwidth]{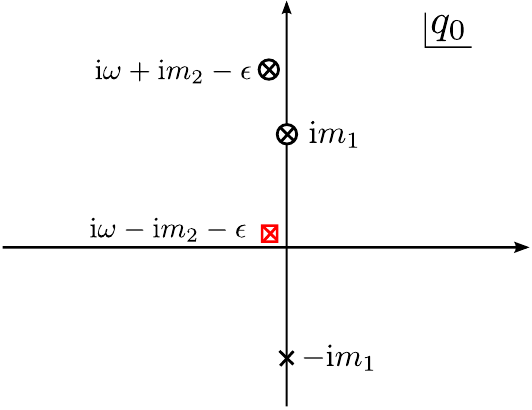}
\caption{Pole structure for the one-loop calculation at $T=0$.}
\label{fig:poles1loop}
\end{figure}
In the present example the difference between the two continuation
procedures  only
consists of residues of poles and no cuts. Explicitly it is given by
\begin{align}\nonumber 
  &\Delta \Gamma^{(2)}_\text{\tiny{res}}(\omega,\vec p;\epsilon)\equiv\Delta
\Gamma^{(2)}_{\text{E,\tiny{cont}}}(\omega,\vec
p;\epsilon)-\Delta\Gamma^{(2)}_{
\text{R}}(\omega,\vec p;\epsilon)\\[2ex]\nonumber
  =&(-\imag) [\Gamma^{(3)}]^2 \int_{\vec q}\sum_\pm \underset{-
    p_0\pm\imag \epsilon_{q+p}^2}{\text{Res}}G_1 G_2
  \Bigg|_{p_0\to -\imag(\omega+\imag \epsilon)}\\[2ex] 
  &\times\left(n(-\omega+\imag\epsilon\mp\epsilon_{q+p}^2)
    -n(\mp\epsilon_{q+p}^2)\right)\,.
\end{align}
In particular, for vanishing temperature, $T\to 0$, where $n(z)\to
-\Theta(-\text{Re}\, z)$, we have
\begin{align}
  n(-\omega+\imag\epsilon\mp\epsilon_{q+p})-n(\mp\epsilon_{q+p})\to
  \pm\Theta(\mp\omega-\epsilon_{q+p})\,.
\end{align}
This implies in particular that at vanishing temperature both results
agree for small Minkowski external momenta $|\omega|<m_2$. This is
simple to understand from the pole structure shown in
\Fig{fig:poles1loop}.  At zero temperature the difference between the
two results is just given by the contribution of the circled pole in
\Fig{fig:poles1loop}, which enters the upper/lower half plane for
$|\omega|>m_2$. However, even at nonvanishing temperature the desired
$\Gamma^{(2)}_\text{R}(\omega,\vec p)$ can be computed entirely numerically
via
\begin{align}
\label{eq:polecorrection}
\Gamma^{(2)}_\text{R}(\omega,\vec p)&=-\lim_{\epsilon\to 0}\left(\Gamma^{(2)
    }_{\text{E,\tiny{cont}}}(\omega,\vec p;\epsilon)-\Delta
  \Gamma^{(2)}_\text{\tiny{res}}(\omega,\vec p;\epsilon)\right)\,.
\end{align}
The calculation of $\Gamma^{(2)}_{\text{E,\tiny{cont}}}$ involves a
straightforward Euclidean integration/ Matsubara sum. The difference
$\Delta \Gamma^{(2)}_\text{\tiny{res}}(\omega,\vec p;\epsilon)$ is
calculable given the residues of the integrand, that can be determined
numerically. The advantage of such a procedure is that given
appropriate regulator functions such as the ones discussed in the next
section, the momentum integration/Matsubara summation can be carried
out as before; it only remains to trace poles of the propagators and
calculate corresponding residues in order to obtain the full real time
result.

\subsection{General-purpose regulators for complex momenta}
\label{sec:RkImagTime}
The previous section gave us a very clear picture of how to treat the
poles of the Euclidean propagator in order to obtain the retarded real
time correlation function. In the present work we aim at evaluating
spectral functions in the framework of the FRG. There, an infrared
cutoff is introduced by means of a momentum- and/or frequency-dependent
modification of the classical kinetic term.  For the simple example of
a real scalar theory this entails
\begin{align}\label{eq:Reg}
  \int_ p \phi(p) p^2 \phi(-p) \to \int_ p \phi(p) \left( p^2+
    R_k(p_0,\vec p)\right)\phi(-p)\,.
\end{align}
For low spatial momenta/frequencies the cutoff function $R_k(p/k\to
0)\approx k^2$ acts as a mass term. In turn, $R_k(p/k\to \infty)\to 0$
ensures that the ultraviolet is unchanged. Lowering the infrared
cutoff scale $k$ leads to a successive integration of the momentum
and/or frequency fluctuations at about the cutoff scale. The scale
dependence of the effective action $\Gamma_k[\phi]$ of the theory at
hand is governed by the Wetterich equation \cite{Wetterich:1992yh},
\begin{align}\label{eq:flow} 
  \partial_t \Gamma_k[\bar\phi]= \012 \Tr \, G[\bar\phi]\,\partial_t R_k\,,
  \quad \quad t=\log k/\Lambda\,,
\end{align}
with the propagator $G[\bar\phi]$ defined in \eq{eq:propcomplex} in 
the presence of a general background. In terms of the effective action it reads 
\begin{align}\label{eq:propG}
G[\bar\phi] = \,\0{1}{\Gamma^{(2)}_k[\bar\phi]+R_k}\,.
\end{align}
In \eq{eq:flow} we have also introduced some reference scale $\Lambda$,
usually being the UV scale, where the flow is initialised. 

In the present work we shall discuss the case of general correlation
functions at the example of the propagator. Its flow, or rather that
of $\Gamma^{(2)}_k$ is derived from \eq{eq:flow} by taking the second
derivative w.r.t.\ $\phi$. The corresponding diagrams are directly
linked to the one-loop diagrams in the last section. This is seen by
rewriting \eq{eq:flow} as 
\begin{align}\label{eq:flowtilde} 
  \partial_t \Gamma_k[\bar\phi]= 
\012 \Tr \,\left. \partial_t\right|_{\Gamma_k^{(n)}}   
\ln \left(\Gamma_k^{(2)}[\bar\phi] +  R_k\right)\,. 
\end{align}
$\phi$-derivatives commute with the partial $t$-derivatives and hence
the flows of $n$-point functions take the form of partial
$t$-derivatives of the corresponding one-loop diagrams with full
propagators and vertices. The partial $t$-derivative only hits the
$R_k$ dependence of the propagators with 
\begin{align}\label{eq:dRk}
  \partial_t|_{\Gamma_k^{(n)}} G = \tilde \partial_{t} G = - G
 \, \dot R_k\, G\,,\quad {\rm with} \quad \dot R_k=\partial_t R_k\,, 
\end{align}
where we have introduced the notation $\tilde \partial_t$ for the
partial $t$-derivatives at fixed $\Gamma_k^{(n)}$. The flow of
correlation functions, $\partial_t \Gamma^{(n)}$, is derived from
\eq{eq:flow} by taking the $n$th derivative w.r.t.\ the fields. With
\begin{align}\label{eq:dphiG}
\0{\delta }{\delta \phi} G=-  G\, \Gamma^{(3)}\, G\,, \quad 
\0{\delta }{\delta \phi} \Gamma^{(m)} = \Gamma^{(m+1)}\,,  
\end{align}   
this leads to a sum of general one loop diagrams in full vertices and
propagators for the flows $\partial_t \Gamma^{(n)}$. However, in each
diagram one of the lines is given by $G \dot R_k G$. As already
discussed in the last section, this allows us to define a unique
analytic continuation procedure in frequency space as follows: In each
diagram contributing to the flow of a given correlation function
$\Gamma_k^{(n)}$ we choose the frequency routing such that $G \dot R_k
G(q)$ has the Euclidean loop frequency $q_0$. This entails that we only
have to discuss the poles of the propagator $G(q+\sum p_i)$ itself as the
frequency argument of $G \dot R_k G(q)$ is Euclidean. It even only takes
values on the Matsubara frequencies. Here $p_i$ are the external
four-momenta with possibly complex frequencies $(p_0)_i$. Note that
the general analysis also includes the poles of the vertices, the
analysis of which will be published elsewhere.

Accordingly, the pole analysis of the propagator poles of the previous 
Section~\ref{sec:poleprocedures} carries over to the present
case. However, the infrared regulator function changes the pole (and
cut) structure discussed above in Section~\ref{sec:poleprocedures}:

First of all, the regulator necessarily triggers cutoff-dependent
shifts of the location of the poles evaluated in
Section~\ref{sec:poleprocedures}. This also extends to possible cuts
in the complex plane, see \cite{Windisch:2012zd,Windisch:2012sz,%
  Windisch:2013mg,Windisch:2013dxa} for an extensive discussion, that
have not been discussed in Section~\ref{sec:poleprocedures}. However,
if these $k$-dependent cuts are present for the one-loop diagrams in
full propagators and vertices, the partial $t$-derivative converts
them into poles. The cuts are reinstated within the $k$-integration
due to the $k$-propagation of the poles. The propagation of the latter
has to be followed anyway, and cuts pose no further problem. Secondly,
in the case of frequency-dependent regulators further, unphysical
poles are generated by the regulator itself.

In summary this leaves us with two options: We avoid the unphysical
regulator poles with a regulator that only depends on spatial
momenta. However, this comes at the price of a breaking Euclidean and
Minkowski symmetry.  The second option is to use space-time
symmetry-preserving regulator functions, that are functions of the
four-momentum squared. Then, the regulator poles cannot be avoided.

\begin{figure}[t]
\centering
\includegraphics[width=\columnwidth]{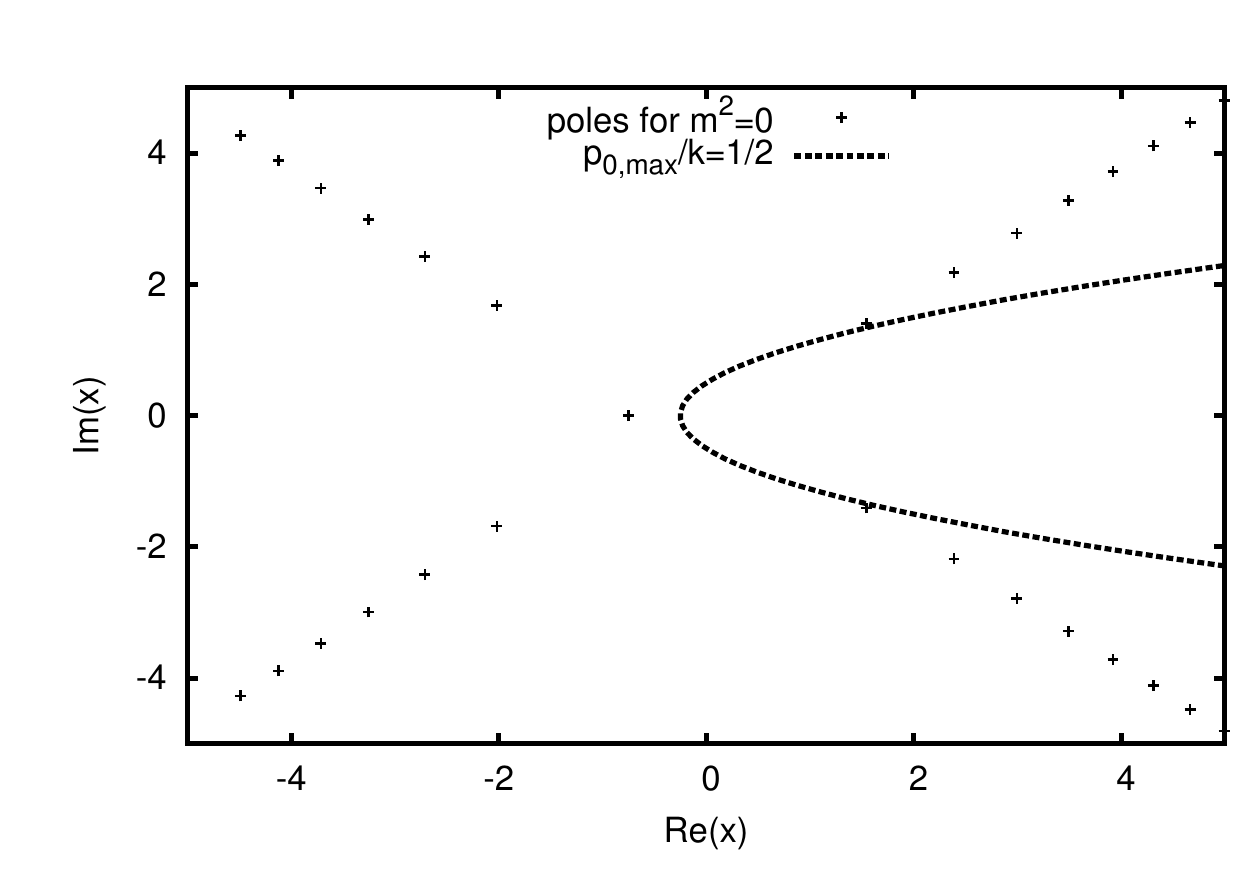}
\caption{Poles of the regularised propagator in the complex
  $x=(p_0^2+\vec p^2)/k^2$-plane for $m^2=0$ and a double
  exponential regulator. The parabola represents the boundary of
  accessible momenta for $p_{0,\text{max}}/k=1/2$.}
\label{fig:poleillustration}
\end{figure}

Here, we are particularly interested in the case of
symmetry-preserving regulator functions in contradistinction to the
predominantly used spatial regulator functions in first applications
towards a direct computation of real time correlation functions.  Such
symmetry-preserving regulator functions are an indispensable
prerequisite for studies of complex systems involving nontrivial
momentum and frequency dependencies, see e.g.\ \cite{Mitter:2014wpa}
for a possible application in QCD. Moreover, most approximation
schemes in use within FRG applications are built on momentum locality:
they do not take into account the full momentum dependence of all
correlation functions at finite order of the expansion scheme. This
applies in particular to the LPA-type expansion used here. Such
expansions asks for regulators that minimise the momentum transfer in
the flow diagrams, see \cite{Pawlowski:2005xe}.

Regulators that minimise the momentum transfer in the diagrams
necessarily need a rapid decay with frequency and spatial
momenta. Such a rapid decay is directly linked to the occurrence of a
large number of unphysical poles, in the case of the numerically
well-tractable exponential decay one has to deal with a whole series
of poles in the complex plane. In \Fig{fig:poleillustration} we show
the propagator \eq{eq:propG} in the presence of a double exponential regulator,
\begin{align}\label{eq:expR}
R_k(p^2)=p^2 r(p^2/k^2)\,,\quad \quad r(x)= \0{x}{e^{x^2}-1}\,, 
\end{align}
and $\Gamma_k^{(2)}\to p^2$ for the sake of simplicity. \Eq{eq:expR}
is a specific case of a general class of regulators with exponential
decay introduced later, see \eq{eq:expregmain}. \Fig{fig:poleillustration} shows an
infinite number of additional regulator poles. Note that in the present massless case the
poles remain fixed in the dimensionless $x=(p_0^2+\vec p^2)/k^2$-plane, for 
the general analysis see App.~\ref{app:deltagconstraints}.

As additional regulator poles at finite $k$ turn out to be
unavoidable, the following observation is helpful: The calculation of
spectral functions $\rho(\omega, \vec p)$ with external frequencies
$|\omega|<p_{0,\text{max}}$ utilises propagators in a strip
$\mathcal{S}_{p_0,\text{max}}$ with
$|\text{Im}\,q_0|<p_{0,\text{max}}$ around the real axis in the
complex $q_0$-plane as input. Therefore one requires regulator
functions which vanish in the strip as $k\to 0$,
\begin{equation}
\label{eq:propcondition}
\lim_{k\to 0} R_k(q+\imag p)\to 0\quad\text{for}\,\, q\in 
\mathbb{R}\,\,\text{and}\, |p_0|<p_{0,\text{max}}\,,
\end{equation}
where here and in the following we use the shorthand notation $q+\imag
p$ for $(q_0+\imag p_0,\vec q+\vec p)$. Note that the most desirable
regulator function would preserve space-time symmetries but show at
most physical poles at finite RG scales $k$. The restriction to
external frequencies in $\mathcal{S}_{p_0,\text{max}}$ restricts the
accessible momenta to a parabola in the complex $x=(p_0^2+\vec
p^2)/k^2$-plane, see \Fig{fig:poleillustration}.

The calculation of Minkowski correlation functions from the Euclidean
ones requires the integration along deformed contours in the complex
$q_0$ plane. Hence, on the one hand we need analytic regulator
functions in order to be able to apply analyticity arguments. On the
other hand we aim for regulator functions which at finite RG-scales
$k$ introduce as few additional poles in the strip as
possible. 

\begin{figure}[t]
\centering
\includegraphics[width=\columnwidth]{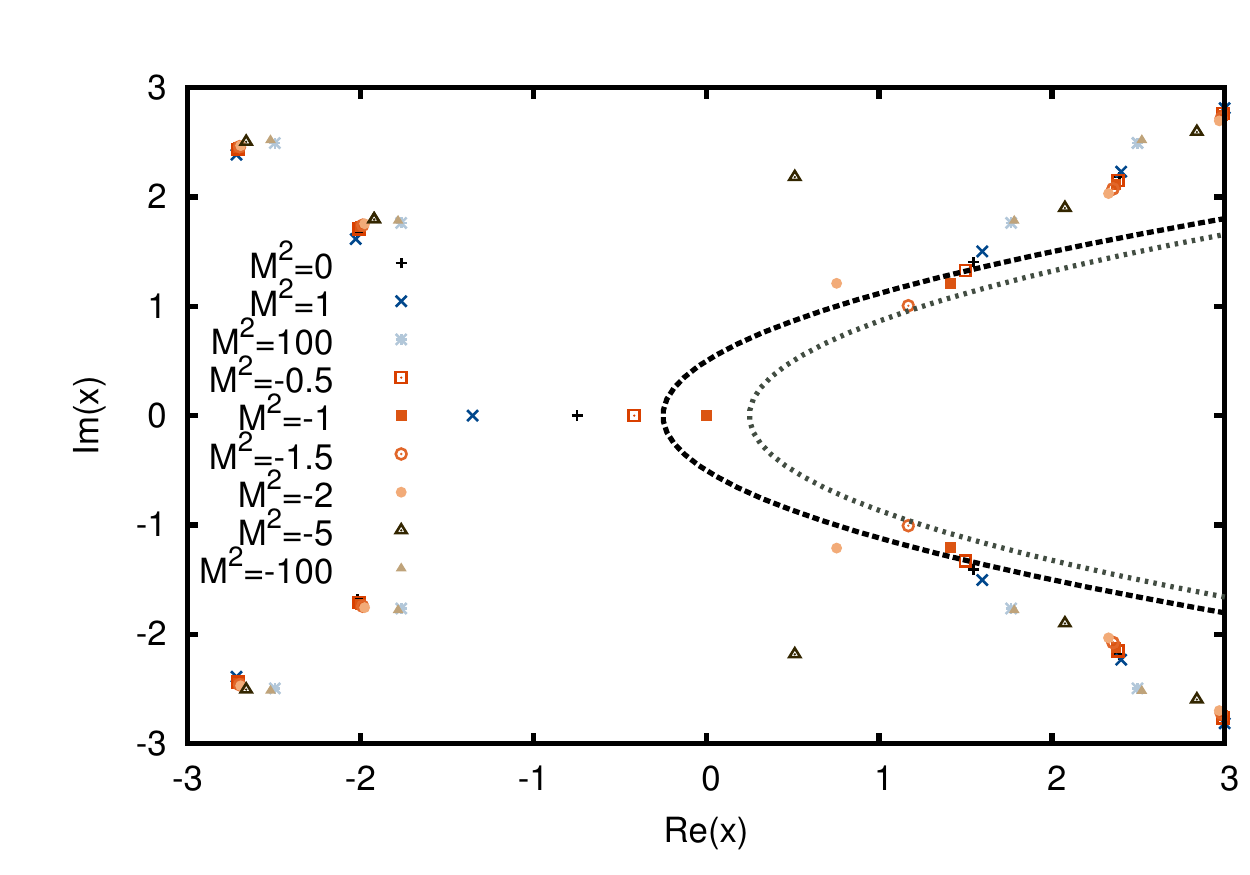}
\caption{Physical (on the negative real axis) and regulator poles in
  the complex $x=(p_0^2+\vec p^2)/k^2$ plane for different values of
  $M^2=(m^2-\Delta m_r^2)/k^2$, see App.~\ref{app:deltagconstraints}
  for details. The black(grey) parabola represents the boundary of
  $\mathcal{S}_{p_0,\text{max}}$ for $\Delta m_r/k=0$ ($\Delta
  m_r/k=1/2$) for $p_{0,\text{max}}/k=1/2$.  }
\label{fig:polelist}
\end{figure}

Here, we construct regulator functions that do not lead to 
any additional regulator poles in addition to the physical poles in the
strip $\mathcal{S}_{p_0,\text{max}}$ at finite $k$. There are various
ways of implementing these constraints, and we have evaluated many of
them. Some of them, including the ensuing symmetry and analyticity
constraints are discussed in App.~\ref{app:deltag} for the benefit of
the reader.

Here we put forward a specific construction that has turned out to be
the most flexible and stable one for our purposes. The central idea is
to introduce an artificial mass term $\Delta m_r^2$ also in the
regulator shape function $r$, which moves the additional regulator
poles outside the strip.  Explicitly, we only discuss the very
convenient class of regulators that are also used in the subsequent
numerical implementation,
\begin{equation}
  \label{eq:reggeneralpurpose}
  R_{k;\Delta m_r^2}(p^2)\! = \!\left(\Delta\Gamma^{(2)}_k(p^2)|_{\phi=\phi_0}
    +\Delta m_r^2\right)\! r\left(\frac{p^2+\Delta m_r^2}{k^2}\right)\,,
\end{equation}
The regulator introduced in \eq{eq:reggeneralpurpose} is proportional
to the momentum-dependent part
$\Delta\Gamma^{(2)}_k(p^2)=\Gamma_k^{(2)}(p^2)-\Gamma_k^{(2)}(0)$ of
the inverse propagator. We want
to stress at this point that the described procedure to avoid artificial regulator poles is of 
general nature and can be applied not only to regulators of the form \eq{eq:reggeneralpurpose} with 
general shape functions but to arbitrary regulators that work for real momenta.

\begin{figure}[t]
\centering
\includegraphics[width=\columnwidth]{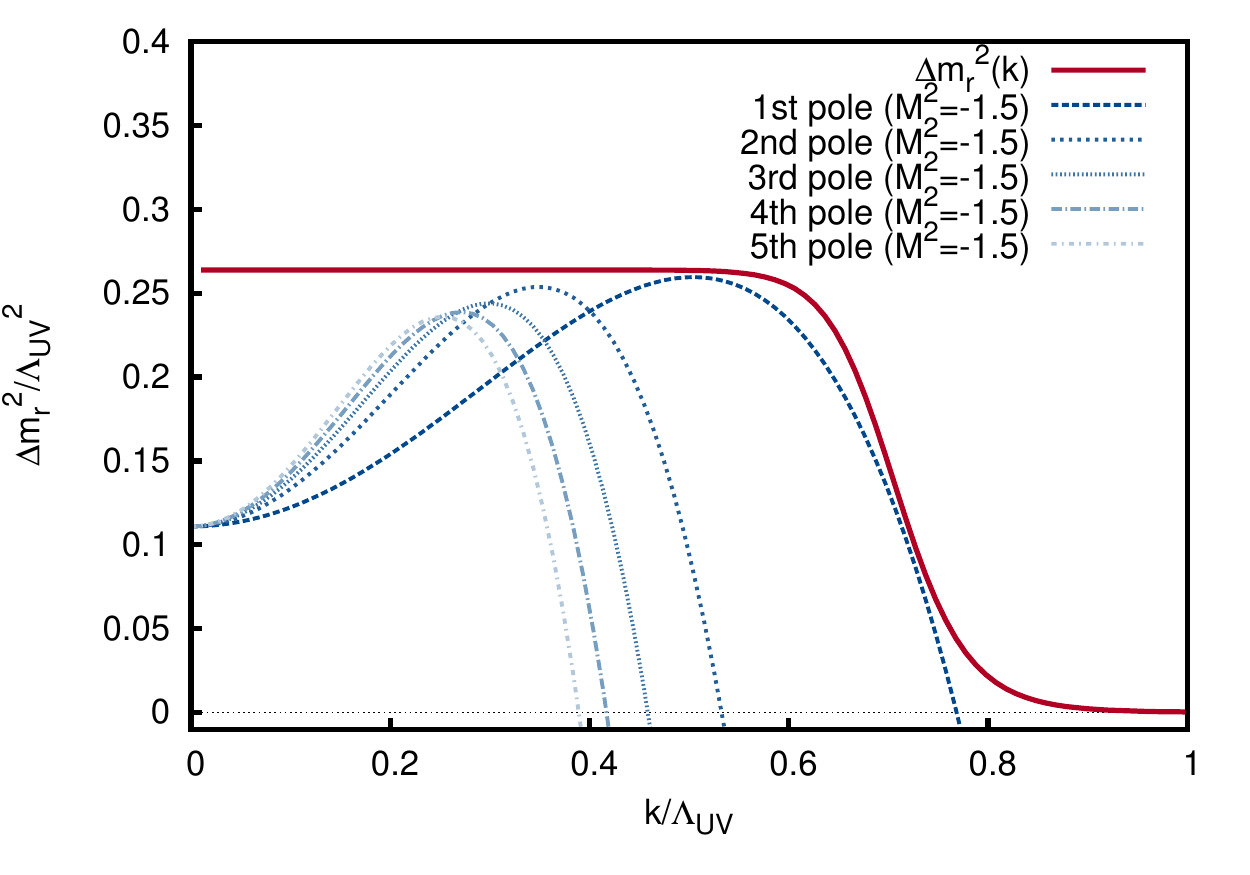}
\caption{Constraints on $\Delta m_r^2(k)$ from avoiding additional
  poles in the parabolic region as obtained from
  \eq{eq:deltagcondition} here for $p_{0,\text{max}}=0.33\,
  \Lambda_{\text{\tiny UV}}$ and a double exponential regulator
  ($m=2$). The solid red line shows our parameterisation
$\Delta m_r^2(k)=\alpha |p_{0,\text{max}}|^2(1+(\beta k/|p_{0,\text{max}}|)^n)^{-1}$
 for $\alpha=2.38$, $\beta=0.47$ and
  $n=20$, see App.~\ref{app:deltagconstraints} for details.}
\label{fig:deltagconstraints}
\end{figure}

\Fig{fig:polelist} shows the pole structure of the propagator in the
presence of a finite $\Delta m_r$-parameter and different mass parameters $m^2$, 
see App.~\ref{app:deltagconstraints} for
a detailed discussion. The main effect of the latter is to shift the
parabolic region of accessible momenta to the right whereas the
location of the poles is only shifted slightly. This analysis implies that
it is always possible to avoid regulator poles in
$\mathcal{S}_{p_0,\text{max}}$ by choosing an appropriately large
$\Delta m_r^2$ at every scale. This implies in particular that one can
start with a vanishing $\Delta m_r^2$ in the UV for $k\to
\Lambda_{\text{\tiny{UV}}}$ with $\Lambda_{\text{\tiny{UV}}} \gg$
physical scales, where the regulator poles are still far outside the
strip.

For our explicit computations we further restrict
\eq{eq:reggeneralpurpose} to a general class of exponential decay
regulators with
\begin{align}
\label{eq:expregmain}
r(x)=\frac{x^{m-1}}{e^{x^m}-1}\,.
\end{align}
\Fig{fig:deltagconstraints} shows
the constraints arising from avoiding the first few propagator poles
in the complex plane for the regulator with $m=2$ in
\eq{eq:expregmain} in units of $\Lambda_{\text{\tiny{UV}}} $. These 
poles can obviously be avoided by a $k$ dependence of $\Delta m_r^2$
in the form of a smooth theta function with appropriately chosen
parameters, see App.~\ref{app:deltagconstraints} for details.

\begin{figure}[t]
\centering
\includegraphics[width=0.95\columnwidth]{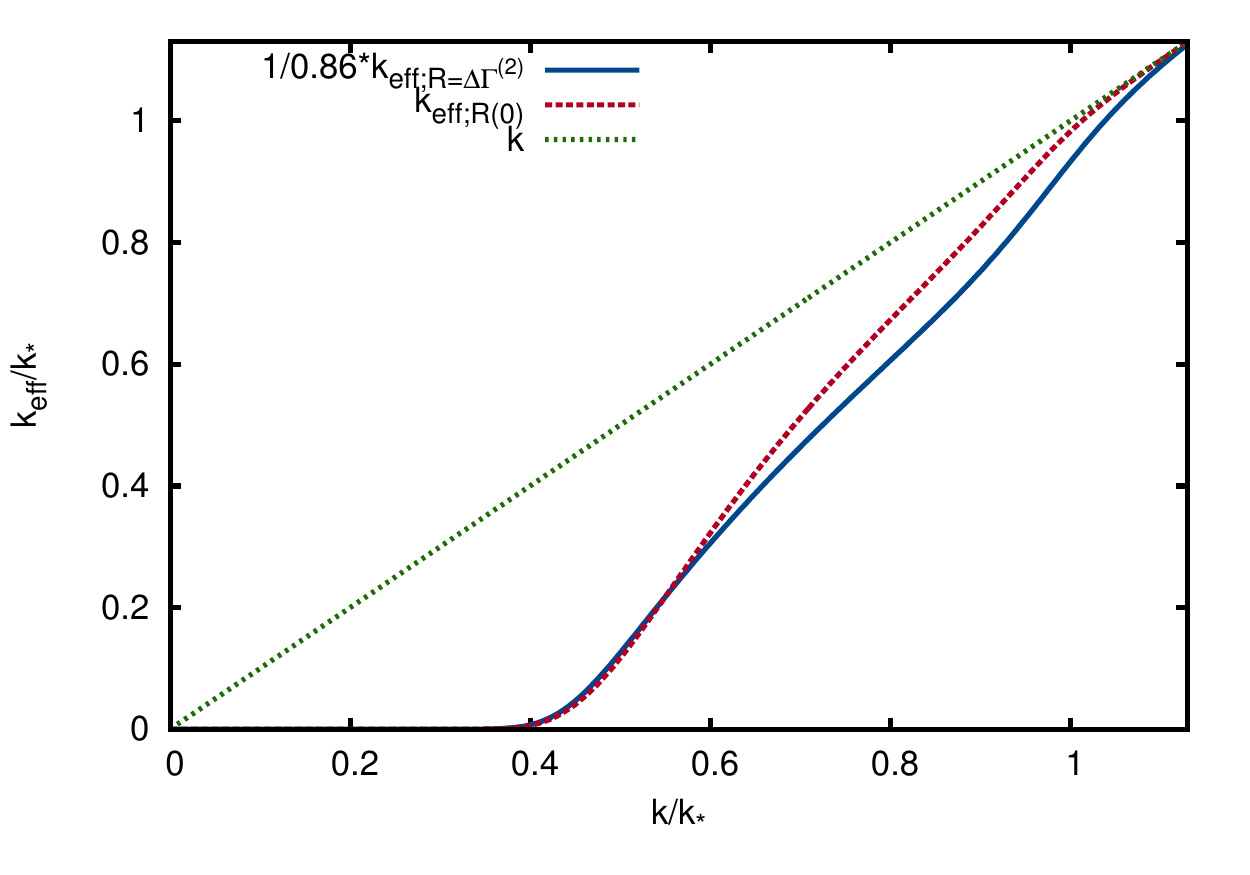}
\caption{Effective cutoff scales $k_{\text{eff};R=\Delta\Gamma^{(2)}}$
  and $k_{\text{eff};R(0)}$ as a function of $k/k_*$, where $\Delta
  m_r^2(k_*)/\Delta m_r^2(0)=0.5$, here for the same parameter set as in \Fig{fig:deltagconstraints}.}
\label{fig:effcutoff}
\end{figure}

The inclusion of $\Delta m_r^2$ leads to a substantial modification of
the regulator's effective cutoff scale \cite{Pawlowski:2005xe,Marhauser:2008fz} compared to the RG-scale
$k$. In particular there is an exponential drop of the physical cutoff
scale $k_{\text{\tiny{eff}}}$ with $\Delta m_r^2$ in the regime $k^2
\lesssim \Delta m_r^2$, see \Fig{fig:effcutoff} and App.~\ref{app:effcutoffscales} for a
detailed discussion. We conclude $\Delta m_r^2(k)$ should be chosen as
small as possible for several reasons.

Firstly we would like to keep the comparability to the standard
regulator case with $\Delta m_r^2=0$, hence allowing for direct access
to the plethora of results obtained there. While this merely is
convenient, we encounter a more severe constraint in theories with
several different field modes with different mass scales. There, the
exponential drop of the physical cutoff scales $k_{\text{\tiny{eff}}}$
relative to the cutoff parameter $k$ causes the momentum fluctuations
of the respective fields to quickly disentangle in the regime $\Delta
m_r^2 \lesssim k^2$ for at least one of the modes. In other words, in
this regime, the flow integrates out momentum and frequency modes of
the different fields at potentially vastly different momentum
scales. This either asks for approximation schemes that are amiable
for a large momentum transfer or a very accurate determination of the
relative physical cutoff scales $k_{\text{\tiny{eff}}}$.

In principle, both properties can be adjusted for. For example, a
vertex expansion keeping the full momentum dependence of the
correlation functions has been put forward in \cite{Mitter:2014wpa}
for QCD. However, the computational effort is relatively large and we
aim at an approach working for general approximation schemes. Appropriately
defined effective cutoff scales can be used to adjust relative cutoff
scales to keep the momentum transfer small in the first place. 
This allows using simpler truncations without considering the full
momentum dependence.

The accurate determination of the relative physical cutoff scale is
the subject of App.~\ref{app:effcutoffscales} and of ongoing work. In
any case, the problem of a potential momentum transfer is minimised by
minimising $\Delta m_r^2$. Therefore, one should for example refrain
from using a divergent $\Delta m_r^2(k)$ as $k\to 0$, that would in
principle allow for arbitrarily large Minkowski frequencies.

\subsection{Application to the $O(N)$ model}\label{sec:ON} 

\begin{figure}[t]
\centering
\includegraphics[width=0.99\columnwidth]{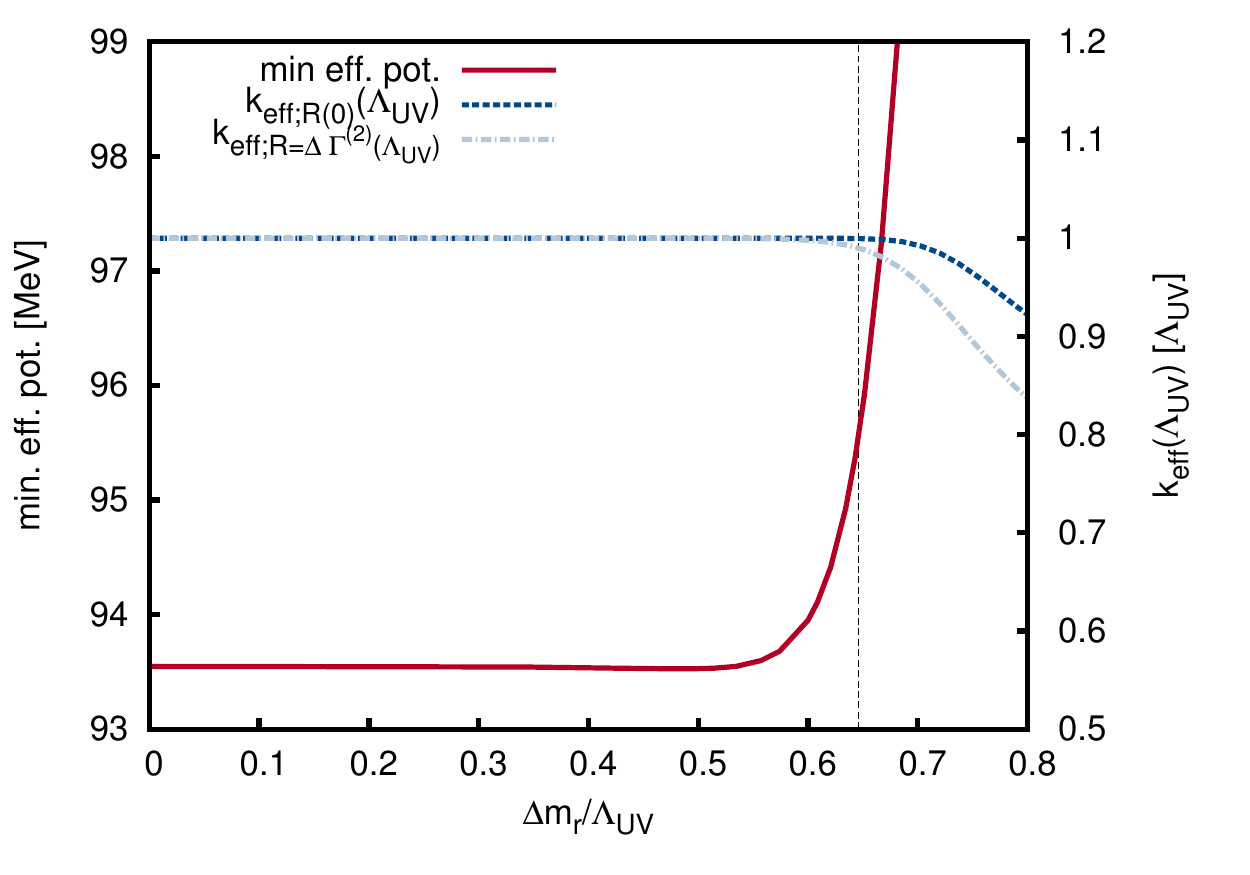}
\caption{Minimum of the effective potential and effective cutoff
  scales $k_{\text{eff}}(\Lambda_{\text{\tiny{UV}}})$ in dependence of
  $\Delta m_r$ obtained by varying $p_{0,\text{max}}$ for fixed
  parameter values $\alpha=2.38$, $\beta=0.47$ and $n=20$ in
  \eq{eq:ansatzdeltag}. The dashed vertical line denotes the value for
  $\Delta m_r/\Lambda_{\text{\tiny{UV}}}$, where
  $k_{\text{eff};R=\Delta\Gamma^{(2)}}(\Lambda_{\text{\tiny{UV}}})$
  deviates more than 1\% from $\Lambda_{\text{\tiny{UV}}}$.  For
  comparison, the value $p_{0,\text{max}}=300$ MeV used in the
  calculation of the spectral functions corresponds to a value $\Delta
  m_r/\Lambda_{\text{\tiny UV}}\approx 0.51$.}
\label{fig:p0indep}
\end{figure}

The approach to FRG computations of correlation functions at complex
frequencies can now readily be applied to general theories including
fermionic fields. No further conceptual or technical problems have to
be solved. Note also that the current set-up is also applicable at
finite chemical potential. In a Euclidean approach the latter
technically is nothing but an imaginary frequency in all correlation
functions. Applications to single spectral functions in QCD and low-energy 
effective models of QCD at vanishing and finite temperature and
density are under way. Apart from providing interesting physics
information about the decay and formation of resonances, they serve as
input in the direct real time computation of transport coefficients
within the approach put forward in
\cite{Haas:2013hpa,Christiansen:2014ypa}, see also \cite{Buballa:2012hb,Lang:2013lla,Lang:2015nca,Ghosh:2015mda}
for recent effective model calculations.
Respective results will be presented elsewhere.

Here we illustrate the power of the present approach for numerical
computations with the calculation of spectral functions in the $O(N)$
model at vanishing temperature in the local potential approximation
(LPA). The $O(N)$-model is a simple model for the mesonic low-energy
dynamics. It is well-known that, despite its formal simplicity, the
solution of the $O(N)$ model in LPA is numerically far more involved
than related quark-meson or NJL-type models in LPA-type
approximations. This is related to the dominance of fermionic short-range fluctuations 
in the latter class of models, as compared to the
dominance of pion long-range fluctuations in the former. Note that in
advanced approximations beyond LPA all models have to face the
numerical subtleties mentioned above. It is this fact which makes the
$O(N)$-model in LPA the ideal test case.

\begin{figure}[t]
\centering
\includegraphics[width=0.99\columnwidth]{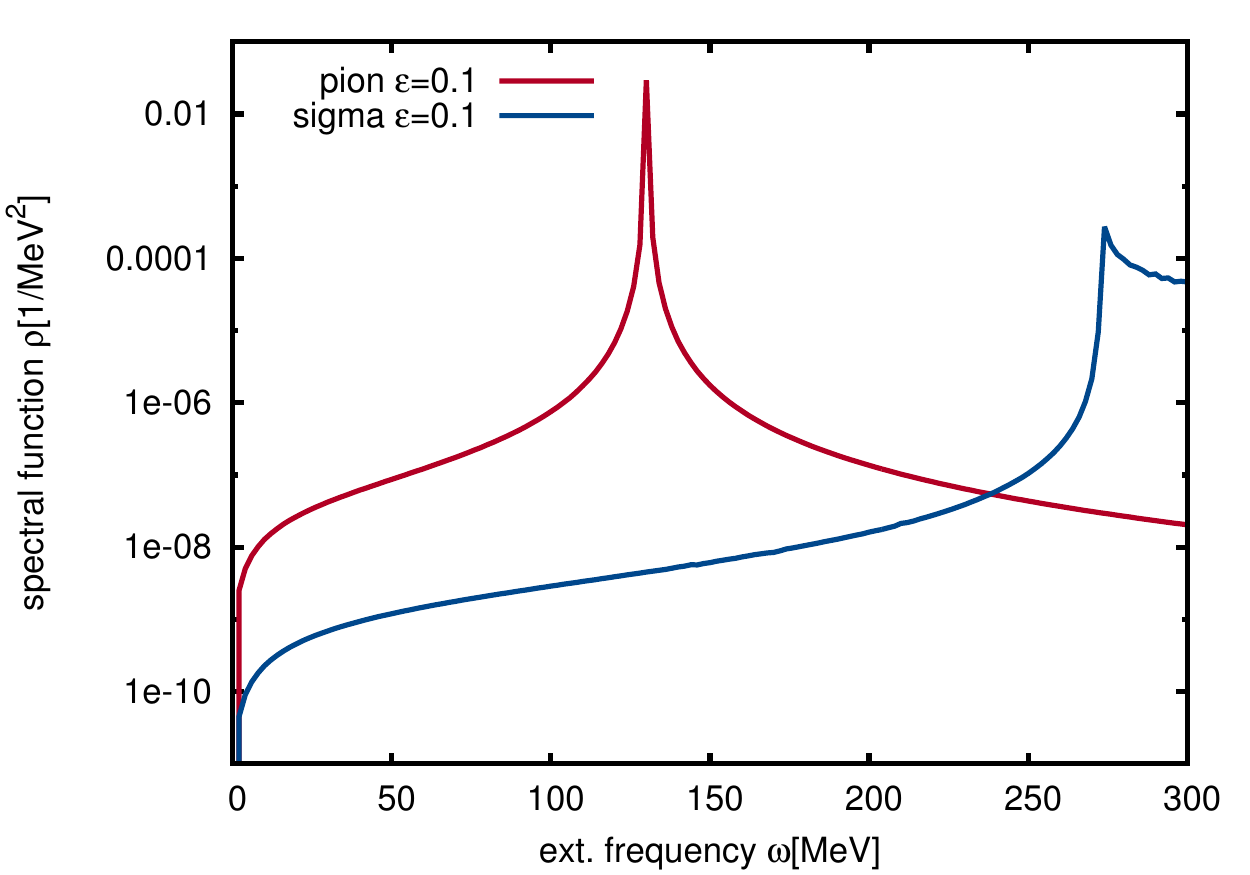}
\caption{Spectral functions of the $O(N)$ model at vanishing
  temperature using a 4d regulator function for $\epsilon=0.1$ MeV.}
\label{fig:spectralon}
\end{figure}

\begin{figure*}
  \centering \subfloat[Pion spectral function.]{
\includegraphics[width=0.48\textwidth]{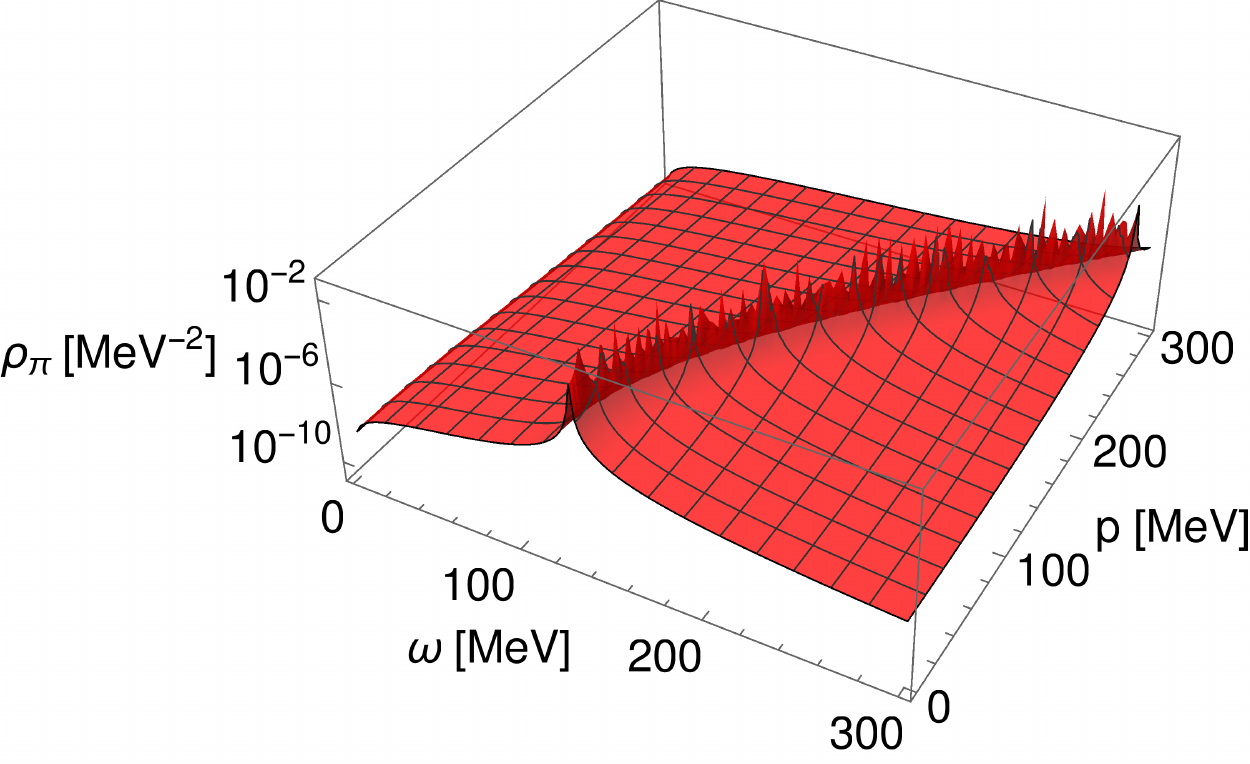}
    \label{fig:specpip}} \hfill \subfloat[Sigma spectral function.]{\includegraphics[width=0.48
\textwidth]{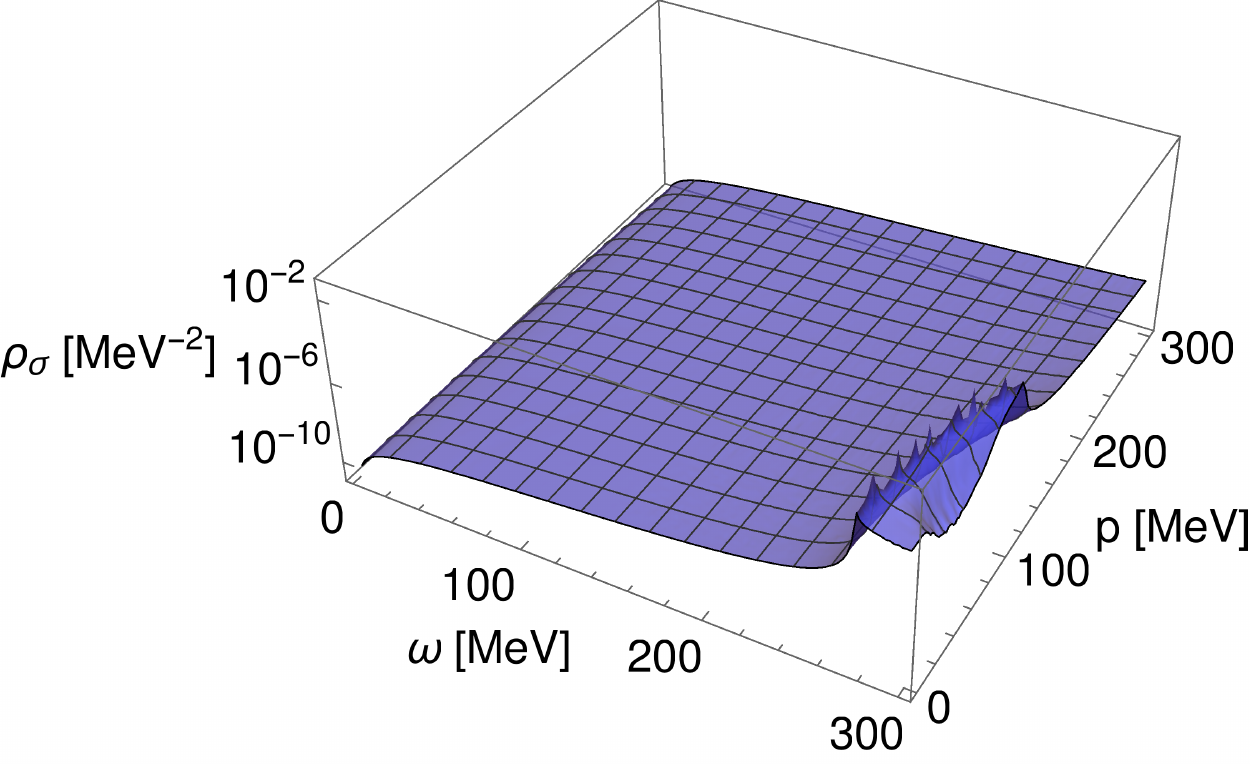}
\label{fig:specsip}}
\caption{Mesonic spectral functions at vanishing temperature as a function of external 
frequency $\omega$ and momentum $|\vec p|$ obtained using a 4d regulator function 
for $\epsilon=0.1$ MeV.\hfill\textcolor{white}{.} }\label{fig:specp}
\end{figure*}
In LPA we take into account the full, scale dependence in the
effective potential, the effective action evaluated at constant
background fields, divided by the four-dimensional volume ${\cal V}$,
\begin{align}\label{eq:effpot}
V_k(\phi) = \0{1}{{\cal V}}\Gamma_k[\phi]\,,\quad {\rm with}\quad 
{\cal V}=\int  \d^4 x \,. 
\end{align}
In the LPA, the effective potential is computed from its flow with the Ansatz 
\begin{align}\label{eq:LPA}
\Gamma_k[\phi]= \int_x \left[ \012 (\partial_x \phi_i)^2 + V_k(\phi^2)\right]\,, 
\end{align}
which gives a closed flow equation for $V_k$. Then the momentum- and
complex-frequency-dependent flow of the propagator is computed with
the vertices $V^{(3)}_k, V^{(4)}_k$ taken from the scale-dependent
effective potential in the LPA approximation, see
App.~\ref{app:deltagflows}. This can be seen as the first iterative
step in the computation of the fully self-consistent computation of
effective potential and propagator as put forward in
\cite{Helmboldt:2014iya}. It has been shown there that the procedure
converges rapidly and that in particular the momentum and frequency
dependence of the propagator is already well approximated by the first
iteration. The implementation follows App.~\ref{app:deltagflows} and
represents a generalisation of \cite{Kamikado:2013sia} to the case of
symmetry-preserving 4d regulator functions. Here we employ a Taylor
expansion at a fixed expansion point \cite{Pawlowski:2014zaa} at a UV
cutoff scale $\Lambda_{\text{\tiny{UV}}}=900$ MeV using an exponential
regulator with $m=2$ with parameters $\alpha=2.38$, $\beta=0.47$ and
$n=20$ in \Eq{eq:thetafn} and map out the pion and the sigma meson
spectral functions up to external real frequencies $\omega \leq
p_{0,\text{max}}=300$ MeV, keeping a small but finite imaginary
frequency $\epsilon=0.1$ MeV. The UV parameters were tuned to yield
physical parameters $f_\pi=93.6$ MeV, $m_\pi=137$ MeV and
$m_\sigma=425$ MeV in the IR.

As a first step we demonstrate the independence of the results of the
chosen maximal value of $\Delta m_r^2$. This is illustrated in
\Fig{fig:p0indep}, where we show the impact of increasing $\Delta
m_r^2$ on the minimum of the effective potential, which was tuned to
physical values for a standard regulator with $\Delta m_r^2=0$. Here
the crucial question is in how far the UV propagator is modified by
the $\Delta m^2_r$-regulator compared to the standard regulator. This
can be quantified by considering effective initial cutoff scales
$\Lambda_{\tiny{\text{eff}}}=k_{\tiny{\text{eff}}}(\Lambda_{\tiny{\text{UV}}})$
as a function $\Delta m_r/\Lambda_{\tiny{\text{UV}}}$. As soon as
$\Lambda_{\tiny{\text{eff}}}$ starts to deviate significantly from
$\Lambda_{\tiny{\text{UV}}}$ the results at $k=0$ are changed. As
shown in \Fig{fig:p0indep} for $\Delta
m_r/\Lambda_{\text{\tiny{UV}}}\gtrsim 0.65$ this deviation exceeds
1\%. The larger deviations in the minimum of the effective potential
beyond this value can be understood on the level of the effective
cutoff scale.  For $\Delta m_r/\Lambda_{\text{\tiny{UV}}}\gtrsim 0.65$
the effective initial scale $\Lambda_{\tiny{\text{eff}}}$ is
lowered. In other words, then we initialise the flow at an effective
lower scale $\Lambda_{\tiny{\text{eff}}}<\Lambda_{\tiny{\text{UV}}}$
with the same amount of symmetry breaking as initialised at
$\Lambda_{\tiny{\text{UV}}}$ for the reference flow at $m_r=0$. On the
other hand, the effective cutoff scale does not capture all effects of
$\Delta m_r^2$ on the momentum-dependent propagator at the UV
scale. For example at $\Delta m_r/\Lambda_{\text{\tiny{UV}}}=0.61$,
the respective propagator deviates by 1\% from its counterpart at
$\Delta m_r^2=0$ in the intermediate momentum regime, whereas the
effective initial scales differ only by 0.3\%.

The expansion of the effective potential about the IR-minimum of the
effective potential similar to the expansion on a grid in field space,
allows a direct computation of the two-point function at the minimum of
the potential where exactly these couplings enter, which avoids having
to expand both the potential and the two-point function about a
scale-dependent minimum \cite{Kamikado:2013sia}. The flow equations
for the two-point functions are then subsequently solved applying
\Eq{eq:polecorrection} by tracking the physical poles of the
propagators in every $k$-step. The resulting spectral function is
shown as a function of frequency in \Fig{fig:spectralon} for
frequencies $\omega+i \epsilon$ and $\epsilon=0.1$ MeV. The clear peak
in the pion spectral function singles out the pole mass of the pion
$m_{\pi, \text{\tiny{pol}}}$, while the sigma shows the threshold of
the $\sigma\to \pi\pi$ decay. Both structures get sharp in the limit
$\epsilon\to 0$, more details and a quantitative comparison to earlier
results obtained using a spatial flat regulator function
\cite{Kamikado:2013sia} are provided in Sec.~\ref{sec:onrealtime}.

In \Fig{fig:spectralon} the spectral functions are evaluated at
vanishing spatial momentum $p=|\vec p|=0$.  In our fully numerical
procedure the extension to finite external spatial momentum represents
only a minor complication as it only requires to evaluate an
additional angular integration. The additional numerical costs are
negligible, and the resulting spectral functions are shown in
Fig.~\ref{fig:specp}. For the pion spectral function the ridge follows
the mass shell relation $\omega^2 =m_{\pi, \text{\tiny{pol}}}^2+\vec
p^2$, while the $\sigma\to \pi\pi$ threshold follows the relation
$\omega^2 =4 m_{\pi}^2+\vec p^2$. The appearance of the curvature mass
$m_{\pi}$ in the threshold relates to the missing completion of the
iteration defined in \cite{Helmboldt:2014iya} that yields the full
frequency and momentum dependence of the propagators. In the fully
iterated result we have $ m_{\pi}^2\to m_{\pi,
  \text{\tiny{pol}}}^2$. For similar results within the quark-meson
model and a spatial regulator function see \cite{Tripolt:2014wra}.

We emphasise that 4d regulators are particularly beneficial for
imaginary time computations with full frequency and momentum
dependencies at finite temperature, see
\cite{Fister:2011uw,Fister:2015eca}. To see this it is illustrative to
consider standard one loop Matsubara sums. These can be performed
analytically and give rise to the well-known thermal suppression
factors $\exp \{-m_{\text{\tiny{gap}}}(p)/(2 T)\}$ in the presence of a
mass gap $m_{\text{\tiny{gap}}}(p)$. This has the additional benefit
of a further UV regularisation of the spatial momentum integrations.
In turn, a numerical computation of the corresponding Matsubara sums
and spatial momentum integrals faces a combined numerical accuracy
problem. Firstly, the exponential decay may involve large explicit
sums of Matsubara frequencies before the spatial integration is
performed. Reversing the order of numerical sum and integration leads
to the requirement of an exponential accuracy of the momentum
integration. In either way the numerical costs rise drastically in
comparison to the vanishing temperature case. This applies in
particular to the interesting transition region of $
m_{\text{\tiny{gap}}}(p) \approx T$.

With 4d regulators with rapid decay the above finite temperature
problem is cured, and we only have to explicitly sum over a relatively
small number of Matsubara frequencies. Related results on finite
temperature spectral functions will be presented elsewhere. This is
particularly relevant when considering the high accuracy requirements
to obtain the imaginary part of the propagator when using a spatial
regulator, which leads to significantly larger numerical costs. 
\section{Spectral function from the real time formalism}\label{sec:SpecRT}
As a complementary second approach we discuss the direct calculation
of spectral functions in a real time formalism
\cite{Chou:1984es,Landsman:1986uw,Das:1997gg,LeBellac:2000}.  Real
time applications of the Functional RG were considered e.g.\ in
\cite{Canet:2003yu,Gezzi:2007PhRvB,Jakobs:2007PhRvL,Gasenzer:2008zz,%
  Berges:2008sr,2009EPJST.168..179S,Gasenzer:2010rq,Canet:2010PhRvL,%
Kloss:2010wj,Canet:2011PhRvE, Kloss:2012PhRvE,Sieberer:2013PhRvL,%
Mesterhazy:2013naa,Sieberer:2014PhRvB,Mathey:2014xxa,2014PhRvB..90h5121R,Mesterhazy:2015uja},
for a recent review see also \cite{Berges:2012ty}. Recent practical
applications use spatial regulator functions
\cite{Mesterhazy:2013naa,Mesterhazy:2015uja}. Here we present the
formalism required for the application of the real time formalism to
the calculation of spectral function as starting point for future
studies. This is not only interesting for the equilibrium case
discussed here but, as mentioned in the introduction, also the
appropriate formalism for applications in nonequilibrium
\cite{Berges:2004yj,Berges:2012ty}.

In order to preserve Lorentz invariance one would like to consider
regulator functions which just depend on the Lorentz scalar 
$p^2=-p_0^2+\vec p^2$. In this case the integrand is just a function
of $p^2$ and we can rewrite the momentum integration as a
three-dimensional integration perpendicular to the mass-shell which is
regularised and a one-dimensional integration along the mass-shell
which is not regularised as the Lorentz-invariant regulator functions
remain constant there. In order to achieve a regularisation on the
mass-shell one is lead to consider regulator functions that break
Lorentz invariance, to wit
\begin{equation} \label{eq:p2p0} R_k(p_0,\vec p)= R_{k}(p^2)
  \left(r_0(p_0^2/k^2)+r_s(\vec p^2/k^2)\right)\,.
\end{equation} 
The regulator in \eq{eq:p2p0} is proportional to a Lorentz-invariant
part $R_{k}(p^2)$, which provides a regularisation of off-shell
fluctuations. A further factor is an energy or spatial momentum cutoff
function which provides a regularisation on the mass
shell. Alternatively one can employ just a Lorentz-invariant regulator
$R_{k}(p^2)$ but then one has to carry out principle value
integrations on the mass shell which requires knowledge about the
position of the poles. Moreover, an accurate treatment of poles close
to the Minkowski axis is required. This will be discussed elsewhere. 

Regulators that also regularise frequencies and allow for
straightforward real time computations, in the sense of no poles close
to the real frequency axis, are necessarily complex, for example
\begin{equation}\label{eq:complexR}  
R_k(p_0,\vec p;a) = (1+ \imag\, a)R_k(p_0,\vec p)\,.
\end{equation} 
The regulators in \eq{eq:complexR} have the additional advantage that
for $a>0$ and $R_k(p_0,\vec p)>0, \forall p_0,\vec p$ they
automatically define Feynman propagators. Such regulators have a very
clear physical interpretation as they introduce a finite width for all
propagators. However, they lead to complex flows and introduce the
necessity of an (additional) fine-tuning in order to remove unwanted
complex parts of the effective action at $k=0$. Note also that they
necessarily introduce additional poles that do not permit for simple
analytic continuations of imaginary time results.

As a specific and simplest example we will focus in the following on
applications involving purely spatial and in particular
spatial flat regulator functions.
In the classification from Eq.~(\ref{eq:p2p0})
spatial regulator functions correspond to the choice $R_k(p^2)\equiv
k^2$, $r_0\equiv 0$ and $r_s(x)=x r(x)$ for some generic shape function $r$. 
Such regulators also do not introduce
additional poles to the complex plane of real and imaginary
frequencies, and hence allow for a straightforward analytic
continuation. For the calculation
of spectral functions in the LPA, the real time procedure can be shown
to be equivalent to the imaginary time procedure, but leads to a
particularly convenient representation for numerical
applications. With regard to the inclusion of general frequency
dependencies the real time formalism is however the more flexible
choice.

\subsection{Closed time path FRG-approach}\label{eq:CPTFRG}
\begin{figure}[t]
\centering
\includegraphics[width=0.95\columnwidth]{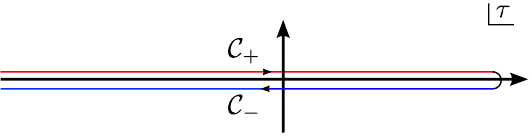}
\caption{Keldysh contour with forward and backward directions. The
  fields $\phi_+$ live on the forward part of the contour $\mathcal{C}_+$, the fields
  $\phi_-$ live on the backward part $\mathcal{C}_-$.}
\label{fig:keldysh}
\end{figure}
The closed time path (CTP) approach is based on the formulation of the
theory on the Keldysh contour, see \Fig{fig:keldysh}. For a scalar
$O(N)$-theory, the field content is then doubled in $\phi_+$ and
$\phi_-$, living on the forward, $\CC_+$, and backward part, $\CC_-$, of the contour
respectively. The action reads
\begin{align}\label{eq:Spm}
  S[\phi_+,\phi_-]=S[\phi_+]-S[\phi_-]\,,\quad S[\phi_\pm]=
  \int_{\CC_\pm} \CL(\phi_\pm)\,,
\end{align}
where $x_\pm$ live on the forward/backward time contour respectively,
$\int_{\CC_\pm}=\int d^4 x_\pm$, and the relative minus sign accounts
for the direction of $x_\pm$ on the contour shown in
\Fig{fig:keldysh}. Naturally, also the regulator term has two parts,
regularising the $\phi_+$ and $\phi_-$ terms,
\begin{align}\label{eq:+-}
  \frac{\imag}{2} \int_{x_+} \phi_+ R_{k}^+ \phi_+ +\frac{\imag}{2}
  \int_{x_-} \phi_- R_{k}^- \phi_- \,.
\end{align}
\Eq{eq:+-} can be generalised to regulators with mixed terms
connecting $\phi_+$ and $\phi_-$. Though potentially advantageous for
technical reasons, such a procedure potentially spoils causality. The
CTP flow equation then takes the form
\cite{Gasenzer:2008zz,Berges:2008sr,Gasenzer:2010rq,Berges:2012ty}
\begin{align}\label{eq:flowkeldysh}
  \partial_t \Gamma_k[\phi_+,\phi_-] = \0{\imag}{2}\Tr\,
  G_{++} \partial_t R_{k}^++ \0{\imag}{2}\Tr\, G_{--} \partial_t
  R_{k}^-\,,
\end{align}
where $\imag G_{++}$ and $\imag G_{--}$ are the $++$ and $--$
components of the propagator. In thermal equilibrium the propagators
can be parameterised solely in terms of the spectral function, see
App.~\ref{app:ctpformalism} and App.~\ref{app:ctpfloweq} for details.

In order to make use of the plethora of explicit resummations and
structural results in perturbative thermal field theory, it is
advantageous to rewrite the flow equation in terms of partial
$t$-derivatives at fixed $\Gamma_k^{(n)}$ of the corresponding one
loop diagrams (with full propagators and vertices) as already done in
Section~\ref{sec:RkImagTime} for the imaginary time formulation.

Moreover, the flow equation \eq{eq:flowkeldysh} includes the flow of
both parts of the contour. Hence it vanishes identically for
$\phi_+=\phi_-=\phi$, $R_k^+=-R_k^-$, and $\pm$-symmetric initial
conditions $\Gamma_\Lambda[\phi_+,\phi_-]
=-\Gamma_\Lambda[\phi_-,\phi_+]$.  A specific subclass of
$\pm$-symmetric initial conditions is given by
\begin{align}\label{eq:initialGkeldysh}
  \Gamma_\Lambda[\phi_+,\phi_-]=\Gamma_\Lambda[\phi_+]
  -\Gamma_\Lambda[\phi_-]\,,
\end{align}
and includes the classical action \eq{eq:Spm}. Taking a $\phi_+$- or
$\phi_-$-derivative breaks the $\pm$-symmetry and gives access to all
correlation functions. This procedure yields for a $\phi_+$-derivative
\begin{equation}
  \partial_{\phi^+}\partial_t \Gamma_k=\0\imag2 \tilde\partial_t 
  \Tr\left(G_{ab}^{\ }\Gamma^{(3)}_{ba+}\right)\,,
\end{equation}
for the canonical choice $R_k^+=-R_k^-=R_k$ and $a,b=\pm$. The
$t$-derivative at fixed $\Gamma_k^{(n)}$, $\tilde\partial_t $, only
hits the propagator, see \eq{eq:dRk}. For the
LPA-based computation of spectral functions as discussed in the
previous Section~ \ref{sec:ON} we have to solve the flow equation for
the effective potential. In analogy to \eq{eq:effpot} it is defined as 
\begin{align}
V_k(\phi_+,\phi_-) =
\0{1}{\cal V}\Gamma_k[\phi_+,\phi_-]\,,
\end{align}
with the space-time volume ${\cal V}$. The LPA approximation at
imaginary time \eq{eq:LPA} trivially extends to the CPT-formulation at
real time as
\begin{align}
  \Gamma_k[\phi]= \sum_{\pm} (\pm)\int_{x_{\pm}} \left[ \012 (\partial_x
    \phi_\pm)^2 + V_k(\phi_\pm)\right]\,, 
\end{align} 
with the single field effective potential $V_k(\phi)$ with
$V_k(\phi_+,\phi_-)=V_k(\phi_+)-V_k(\phi_-)$.  Then, the flow equation
for the effective potential $V_k(\phi)$ reads, 
\begin{align}
  \partial_\phi  \dot V_k(\phi) =&  \left.\partial_{\phi_+} 
    \dot V_k(\phi_+,\phi_-)\right|_{\phi_\pm=\phi}\nonumber\\[2ex]
=& -\012 \tilde \partial_t  \Tr
  \left(\text{Im}\,G^{\ }_{++}\Gamma^{(3)}_{+++}\right)\,.
\end{align}
In the LPA the spectral function is given by a delta function. In the
simplest case of a spatial regulator function, i.e.\ for
$R_k(p^2)\equiv k^2$, $r_0\equiv 0$ and $r_s(x)=x r(x)$ in
\eq{eq:p2p0}, which we will consider in the following, the frequency
integration can be trivially performed. As expected, the flow
equation for a spatial regulator function in the CTP formalism
coincides with the flow equation for the effective potential derived
in the imaginary time formalism, see App.~\ref{app:ctpeffpot} for
details.

To determine the spectral function or correspondingly real and
imaginary parts of the retarded propagator, it is sufficient to derive
flow equations for particular components, $\text{Re}\,
\Gamma^{(2)}_{++}$ and $\Gamma^{(2)}_{+-}$, of the inverse propagator,
see App.~\ref{app:ctpformalism} for details. Again we consider the
$O(N)$ model as simplest application where the flow equations for the
real and imaginary parts of the retarded two-point functions are given
explicitly by \eq{eq:floweqspectralrealtimeon} together with
\eq{eq:relgammabargammar}.  For the case of momentum-independent
vertices, these equations are formulated in terms of
$\tilde t$-derivatives of the loop integrals
\begin{align}
  \bar
  J_{ji}^{\text{Re}}(p)=&\frac{1}{2}\int_q\text{Im}\,G^j_{++}(q+p)\,
  \text{Re}\,G^i_{++}(q)\nonumber\\
  &+\frac{1}{2}\int_q\text{Re}\,G^j_{++}(q+p)\,\text{Im}\,
  G^i_{++}(q)\nonumber\\[2ex]
  \bar J^{\text{Im}}_{ji}(p)=&-\frac{1}{2}\frac{\sign(p^0)}{2
    n(p_0)}\int_qG^j_{+-}(q+p)G^i_{-+}(q)\,.
\end{align}
In the LPA approximation, i.e.\ inserting delta functions for the
spectral functions in the involved propagators, these take the form of
the well-known one-loop expressions in the real time formalism
\cite{LeBellac:2000} and are given explicitly in \eq{eq:Jijre} and \eq{eq:Jijim}.

As discussed in detail in App.~\ref{app:ctponmodel}, the simplest way
towards deriving explicit flow equations for a spatial flat regulator
function is to insert the regulator at this point and to evaluate the
$\tilde t$-derivative afterwards by exploiting the fact that
it only acts on the explicit $k$ dependence in $\epsilon^i_q$ but not
on mass term. In this way one derives explicit flow equations, which
have a particularly simple structure for the imaginary part, see \eq{eq:partialtJ}. The
corresponding flow equation shows delta functions in $k$. Therefore
no numerical integration has to be carried out to evaluate the flow,
see App.~\ref{app:ctponmodel}, which is particularly
convenient for numerical purposes. 

Despite its different appearance
the flow equations for the retarded correlation functions can be
shown to be formally equivalent to those derived starting from an
imaginary time formalism
\cite{Strodthoff:2011tz,Kamikado:2012bt,Kamikado:2013sia,%
  Tripolt:2013jra,Tripolt:2014wra}, see
App.~\ref{app:equivalence}. This reflects nothing but the equivalence
of real time and imaginary time formalisms for the specific example of
the evaluation of one-loop diagrams
\cite{LeBellac:2000,Aurenche:1991hi}. Here the real time formalism lead to
a representation where the $\epsilon\to 0$ limit in the transition
from the Euclidean two-point function evaluated at complex external
momentum and the retarded two-point function, see
\eq{eq:EuclideanRetardedGamma2}, is already taken.  Such a formulation
avoids high numerical costs for the evaluation of the flow equation in
particular at finite temperature and density, see the discussion from
above. From our point of view the real time approach is the preferable
formalism to use for applications with spatial regulator functions, in
particular with regard to the inclusion of further frequency
dependencies in the flow. 

We close with the remark that the above strategy can also be applied
to covariant regulators. They have the advantage of full Lorentz
invariance but additional care has to be taken concerning regulator
poles close to the Minkowski axis as well as the numerical
investigation of principle value integrals. While the former problem
is resolved within the class of regulators defined in
Sec.~\ref{sec:RkImagTime}, the practical implementation of the
latter requires special attention and will be discussed elsewhere.

\subsection{Application to the $O(N)$ model}
\label{sec:onrealtime}

\begin{figure}[t]
\centering
\includegraphics[width=0.99\columnwidth]{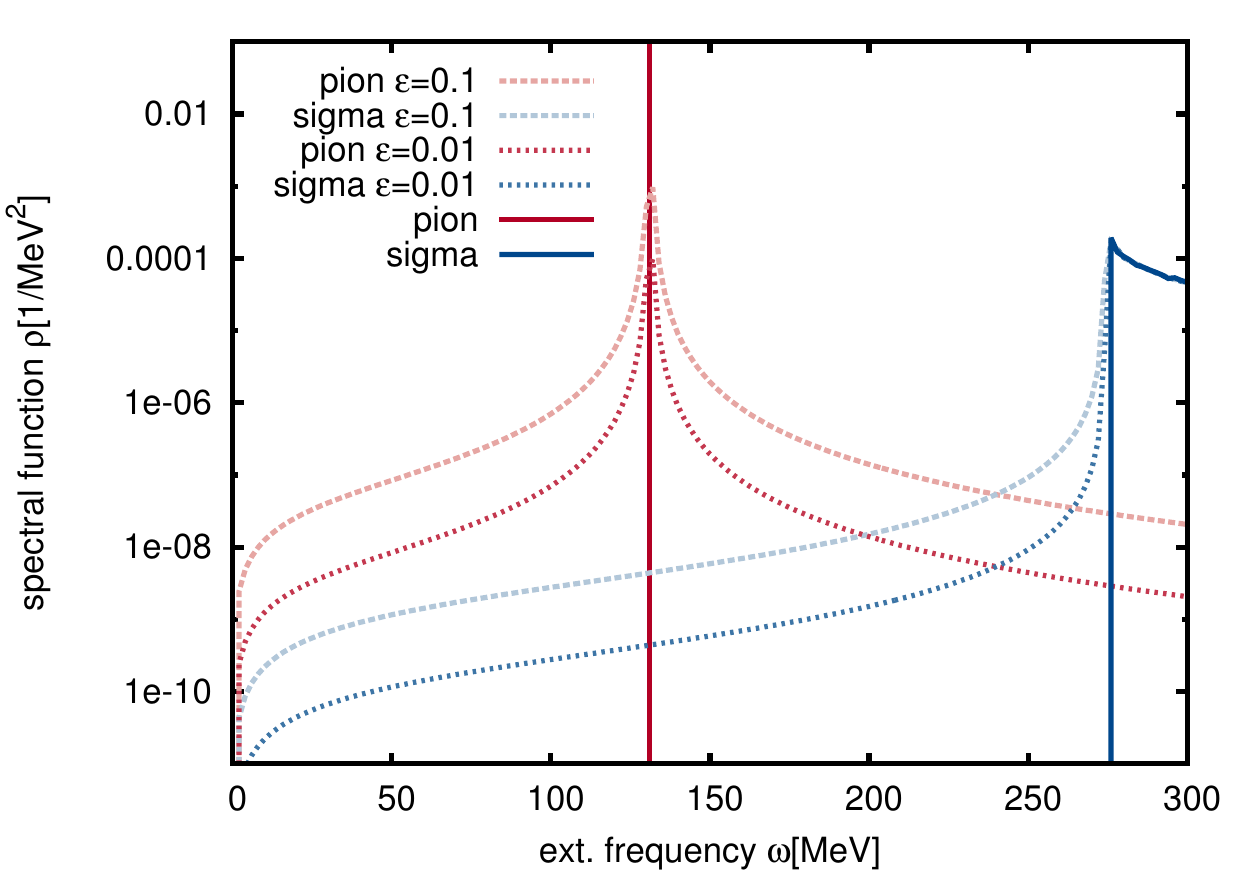}
\caption{Spectral functions of the $O(N)$ model at vanishing
  temperature comparing the result from the real time formalism to
an implementation keeping finite imaginary parts $\epsilon=0.1$ MeV and $\epsilon=0.01$ MeV 
in the external momentum.}
\label{fig:spectralon3d}
\end{figure}

\Fig{fig:spectralon3d} shows the numerical result obtained by
integrating the flow equations \ref{eq:floweqspectralrealtimeon} using
a set-up similar to the one described in
App.~\ref{app:deltagflows}. To ensure comparability of both results
the UV parameters were chosen to yield the same physical parameters in
the IR as in Sec.~\ref{sec:ON}.  In particular we compare the result
from keeping a finite imaginary part in the external momentum,
implemented in the same way as in Sec.~\ref{sec:ON} and in the
original works \cite{Kamikado:2013sia,Tripolt:2013jra}, to the real
time result where the limit $\epsilon\to 0$ has already implicitly
been taken. As clearly visible from the $\sigma$-meson spectral
function at large momenta both results agree once a genuine imaginary
part in the retarded two-point function builds up. The finite imaginary
external momentum $\epsilon$ leads to an artificial broadening of the
spectral functions below the thresholds where the imaginary part has
to vanish identically at zero temperature. In this case the
calculations can even be corrected by hand in order to obtain the full
result with $\epsilon$ extrapolated to zero. The pion pole mass
corresponding to the peak in the pion spectral function agrees in all
cases as it is determined by the zero of the real part of the inverse
retarded correlator which depends only weakly on $\epsilon$.

\begin{figure}[t]
\centering
\includegraphics[width=0.99\columnwidth]{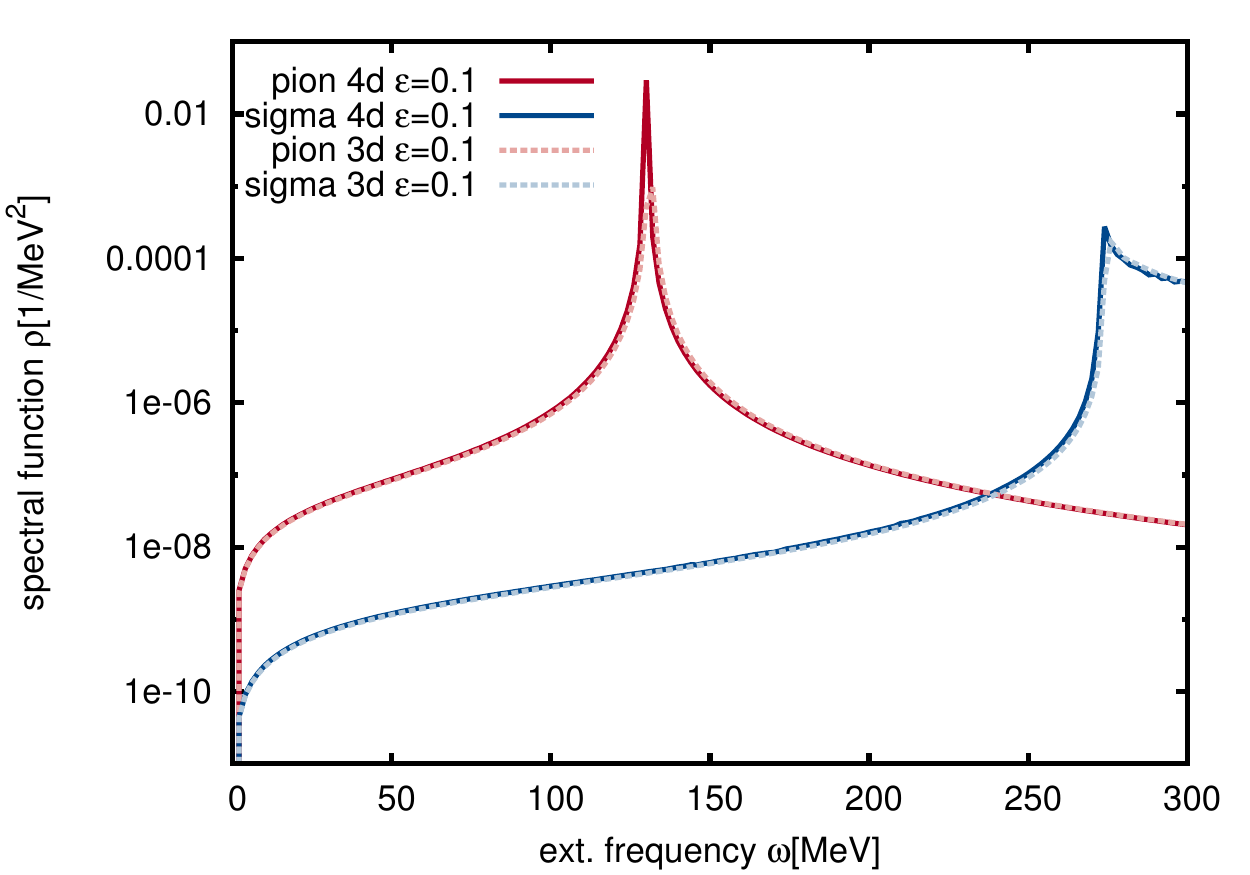}
\caption{Spectral functions of the $O(N)$ model at vanishing
  temperature comparing the result using a spatial 3d regulator and a
  4d exponential regulator in both cases for a finite external
  momentum $\epsilon=0.1$ MeV.}
\label{fig:spectraloncomp}
\end{figure}

Coming to a comparison of the spatial regulator results from this
section to the calculation using a 4d regulator from
Sec.~\ref{sec:ON}, the most meaningful comparison comparison is that
of two calculations with a fixed imaginary external frequency
$\epsilon=0.1$ MeV keeping in mind the effects of the extrapolation
$\epsilon\to 0$ as discussed above. Fig.~\ref{fig:spectraloncomp}
shows a striking agreement between the spectral functions from both
calculations which is a highly nontrivial statement. In both cases the
pion pole mass is given by $m_\pi^\text{pole}=131$ MeV compared to a
curvature mass of $m_\pi=137$ MeV. This difference of 5\% in the LPA
was shown to decrease with the inclusion of the full momentum
dependence of the propagators, see \cite{Helmboldt:2014iya}.

\section{Summary and Conclusions}
In this work we put forward different ways for a direct computation of
real time correlation functions in the framework of the FRG.  Such
approaches avoid the large systematic errors of carrying out a numerical 
analytic continuation of given Euclidean data by a direct
computation of two-point correlation functions for Minkowski external
momenta.

In the first part of the work we put forward a general FRG-framework
that allows to compute Euclidean correlation functions at complex
momenta for general regulators. Then, retarded correlation function
can be obtained from these correlation functions in a fully numerical
procedure. In particular, this involved the construction of general
space-time symmetry preserving regulator functions, which avoid the
occurrence of artificial regulator poles in the momentum region of
interest. Such regulator functions have a broad range of possible
applications not only for real time applications but also for the
closely related problem of the inclusion of a finite chemical
potential with 4d regulator functions. They also allow to use the
plethora of existing Euclidean results for real time physics. The
procedure was put into practice at the example of the computation of
spectral functions in the $O(N)$ model.

The second part of the work was devoted to a complementary approach
starting directly within the real time formalism on the Keldysh
contour. Here we focused on the formalism for the calculation of
spectral functions in the closed time path formalism. Note that for a
spatial flat regulator function the resulting flow equations formally
agree with those derived in the imaginary time formalism. We
confirmed this formal equivalence, the real time representation of the
flow, however, was shown to be particularly amiable towards numerical
applications. The discussion focused on the formalism and applications
were restricted to the $O(N)$ model as a simple, illustrative
example. The extension towards general theories, and in particular the
inclusion of quarks is straightforward. Furthermore, the formalism put
forward in this work is directly applicable to nontrivial frequency
dependencies. 

\acknowledgments 
\hfill\\
\noindent \emph{Acknowledgments} We thank R.~Alkofer, N.~Christiansen, W.-J.~Fu
and L.~von Smekal for discussions. This work is supported by the
Helmholtz Alliance HA216/EMMI and the grant ERC-AdG-290623.
\appendix

\section{General-purpose regulators for complex momenta}
\label{app:deltag}
We consider regulators of the form
\begin{equation}
\label{eq:deltagreg}
R_{k;\Delta m_r^2}(p^2)=\Gamma^{(2)}_k(p^2)|_{\phi=\phi_0}r
\left(\frac{p^2+\Delta m_r^2}{k^2}\right)\,,
\end{equation}
with shape function $r$, which are proportional to $\Gamma_k^{(2)}$ at
some specific field value $\phi_0$.  Hence at least at this particular
field value $\phi_0$ regularised inverse propagator
$\Gamma_k^{(2)}+R_k|_{\phi_0}$ shares the zeros of $\Gamma_k^{(2)}$.
Such regulators are obviously well suited to be used in connection
with Taylor expansions (at $\phi_0$). In this paper we will consider
exponential regulator functions as given in \eq{eq:expregmain}.

The above regulator functions are perfectly suited for applications
where the squared mass parameter stays positive, e.g.\ for a Taylor
expansion at the scale-dependent minimum. However, both in grid
implementations as well as for Taylor expansions at a fixed expansion
point \cite{Pawlowski:2014zaa} the requirement of a positive squared
mass parameter is not always satisfied. This is obviously problematic
for regulators of the form \eq{eq:deltagreg} as it leads to a pole in
$G$. Therefore we also consider slight generalisations of
Eq.~(\ref{eq:deltagreg}) i.e.\ modified regulators of the form
\begin{equation}
\label{eq:reggeneralpurposeapp}
R_{k;\Delta m_r^2}(p^2)\! =\! \left(\Delta\Gamma^{(2)}_k(p^2)|_{\phi=\phi_0}\!
  +\!\Delta m_r^2\right)\! r\left(\frac{p^2+\Delta m_r^2}{k^2}\right)\,,
\end{equation}
that depend only on the momentum-dependent part,
$\Delta\Gamma^{(2)}_k(p)=\Gamma^{(2)}_k(p)-\Gamma^{(2)}_k(0)$, to
circumvent problems for $m^2<0$, and introduces an additional $\Delta
m_r^2\cdot r$-term which avoids the dropping of the effective cutoff
scale to zero for $\Delta m_r^2\gg k^2$, see
App.~\ref{app:effcutoffscales}. This choice of regulator obviously
implies that even at $\phi_0$ the physical poles of $G$ and those of
the regularised propagator $(\Gamma^{(2)}_k+R_{k;\Delta
  m_r^2(k)})^{-1}$ no longer coincide. This is however not an
important requirement as long as we assure that no regulator poles
occur inside $\mathcal{S}_{p_0,\text{max}}$.

Independent of the chosen implementation, the overall picture is the
same in both cases. One introduces an additional mass term in the
regulator shape function in order to shift the additional regulator
poles outside the momentum region of interest, here given by
$\mathcal{S}_{p_0,\text{max}}$. For large RG-scales these will be
outside the strip but it requires a finite $\Delta m_r^2$ at least
towards the IR to ensure that this remains the case during the full
RG-flow. The detailed discussion of the constraints on the $\Delta
m_r^2$-term can be found in App.~\ref{app:deltagconstraints}.

\subsection{Constant $\Delta m_r^2$}
\label{app:constantdeltag}
The simplest way of implementing the pole constraints discussed in the previous section is to use a sufficiently large
constant $\Delta m_r^2$.  As the $\Delta m_r^2$-term is already
present at the UV cutoff scale $\Lambda_{\text{UV}}$ one has to take
into account the change in the initial conditions in the UV in order
to directly compare to results with $\Delta m_r^2=0$. In the following
we describe two procedures for obtaining two-point functions at real
time momenta starting from given initial conditions at $\Delta
m_r^2=0$.  The first possibility is to use a two-step procedure: Given
initial conditions at $\Lambda$ for $\Delta m^2_r=0$, the proper initial
condition for $\Delta m^2_r \neq 0$ is obtained by integrating a flow
equation in $\Delta m_r^2$,
\begin{equation}\label{eq:deltamint}
\begin{split}
  &\Gamma^{(2)}_{\Lambda;\Delta m_r^2} (q+\imag
  p)=\Gamma^{(2)}_{\Lambda;0}(q+\imag p)
  \\[2ex]
  &+\frac{1}{2}\frac{\delta^2}{(\delta \phi)^2}\left(\int_0^{\Delta
      m_r^2} \hspace{-0.7cm}\d \Delta m_r^2{}' (\partial_{\Delta
      m_r^2{}'}R_{\Lambda;\Delta m_r^2{}'}) G_{\Lambda;\Delta
      m_r^2{}'}\right)\!\!(q\!+\!\imag p)\,,
\end{split}
\end{equation}
where the first term is obtained by a trivial continuation of
$\Gamma^{(2)}_{\Lambda;0}$. This procedure has to be carried out for
the effective potential and the two-point function for every value of
$p_0$. Note that for the two-point function the diagrams in \eq{eq:deltamint}
have to be evaluated with the same frequency routing as the flow
itself: $G \partial_{\Delta m_r^2{}'}R_{\Lambda;\Delta m_r^2{}'}
G(q)$ only depends on the loop frequency. As a second step, having
integrated this equation to sufficiently large $\Delta m^2_r\sim
p^2_{0,\text{max}}$, one can now solve the flow equation in $k$,
\begin{equation}
\label{eq:deltagdirectkint}
\Gamma^{(2)}_{k;\Delta m_r^2}(p)=\Gamma^{(2)}_{\Lambda;\Delta m_r^2}
+\int_{\Lambda}^0 d k\,\, \partial_k \Gamma^{(2)}_{k,\Delta m_r^2}\,.
\end{equation}
As an alternative to the direct integration in
\eq{eq:deltagdirectkint} one can make use of an equation for
$\partial_p \Gamma^{(2)}_{k,\Delta m_r^2}(p)$ in order to successively
enter the complex plane starting from the real axis. Therefore we
assume a momentum-independent $\Delta m_r^2$, $\partial_q\Delta
m_r^2=0$. Then we can write the flow equation for
$\partial_p\Gamma^{(2)}_{k;\Delta m_r^2}(p)$ as
\begin{align}
\label{eq:dpgamma2}
  \partial_t\partial_p \Gamma^{(2)}_{k}(p)&=\text{Tr}_q\, G \partial_t R 
  G (q) \partial_p G(q+\imag p)\nonumber\\[2ex]
  &=-\text{Tr}_q\, G\,\partial_t R\,G(q) [G(\partial_p G^{-1}) G](q+\imag p)\,,
\end{align}
where we suppressed the dependence on $\Delta m_r^2$ in order to simplify the
notation. At a given step $p$, this evolution equation involves only
known objects, namely $(G\,\partial_t R\,G)(q)$, $\partial_p G^{-1}(q+\imag
p)=(\partial_p \Gamma^{(2)}_k+\partial_p R)(q+\imag p)$ and $G(q+\imag
p)$.  This is a very clear demonstration of how one can gradually
enter the complex momentum plane starting from data on the real axis.

\subsection{Scale-dependent $\Delta m_r^2(k)$}
\label{app:scaledepdeltag}
The second approach towards satisfying the pole constraints is to introduce a scale-dependent
$\Delta m_r^2$-term which only sets in later during the RG flow. At
the same time this circumvents the necessity of calculating modified
initial conditions in order to connect to calculations at $\Delta
m_r^2=0$ as the first part of two-step procedure introduced in the
previous section as the $\Delta m_r^2$-term vanishes at the UV cutoff
scale. An appropriate scale-dependent $\Delta m_r^2$-term can be
modeled by means of a smooth theta function, i.e.\ using the Ansatz
\begin{equation}
\label{eq:ansatzdeltag}
\Delta m_r^2(k)=|p_{0,\text{max}}|^2 \theta_{\alpha,\beta}(|p_{0,\text{max}}|/k),
\end{equation}
where 
\begin{equation}
\label{eq:thetafn}
\theta_{\alpha,\beta}(x)=\frac{\alpha}{1+\left(\frac{\beta}{x}\right)^n}\,,
\end{equation}
denotes a smooth approximation to a step function with appropriately
chosen constants $\alpha$, $\beta$ and $n\in\mathbb{N}$, see
App.~\ref{app:deltagconstraints} for further details.
\subsection{Pole constraints on $\Delta m^2_r$}
\label{app:deltagconstraints}
In this appendix we analyse the poles of the regularised propagator
$G(p^2)=(\Gamma^{(2)}_k(p^2)+R_k)^{-1}$ in more detail.  In the
following we consider regulators of the general form \eq{eq:reggeneralpurposeapp}. 
A similar analysis goes through for regulators
of the form \eq{eq:deltagreg}.

We are interested in the poles of the regularised propagator or
equivalently in the zeros of
\begin{equation}
(p^2+\Delta m_r^2)(1+r)-\Delta m_r^2+m^2=0\,,
\end{equation}
which reduces for the exponential shape function \eq{eq:expregmain} and
$x\neq 0$ to the condition
\begin{equation}
\label{eq:condition}
\left(e^{x^m}-1\right)\left(x+M^2\right)+x^m=0\,,
\end{equation}
for $x=(p^2+\Delta m_r^2)/k^2$ and $M^2=(m^2-\Delta m_r^2)/k^2$.
It remains to analyse the solutions of Eq.~\ref{eq:condition} for
fixed $M^2$. Leaving aside the case $m=1$, for $M^2>-1$ the equation
admits a negative solution and for $M^2<-1$ two purely imaginary
complex conjugate physical solutions $x_\text{phys}$. In addition
there is an infinite sequence of poles introduced by the regulator. As
shown exemplarily for the double exponential regulator $m=2$ in
\Fig{fig:polelist} for fixed finite $M^2$ these can be determined
numerically. For asymptotically large values $M^2\to\pm\infty$ they
converge towards the poles of the function $x(1+r(x))$ i.e.\ where
$x^m=2\pi \imag N$ for $N\in\mathbbm{N}$.  This gives us access the
solutions $x_{\text{reg}}$ of \eq{eq:condition}.  On the other hand
the restriction of the external momenta to the domain
$\mathcal{S}_{p_{0,\text{max}}}$, i.e.\ $p=(p_{0,R}+\imag p_{0,I},\vec
p)$ for $|p_{0,I}|<p_{0,\text{max}}$ and $p_{0,R}\in\mathbbm{R}$,
corresponds to a parabola in the complex $x$-plane bounded by
\begin{equation}
\label{eq:parabola}
\{t^2-p_{0,\text{max}}^2+\vec p^2+\Delta m_r^2+ 2\imag\, 
t\, p_{0,\text{max}}\,|\,t\in\mathbb{R}\}\,.
\end{equation}
Here we already included a $\Delta m_r^2$-term which adds to
the real part and shifts the parabola in the direction of the positive
real axis. Given a regulator pole at $x_{\text{reg}}=x_R+\imag x_I$ it
is then simple to evaluate the condition
\begin{equation}
\label{eq:deltagcondition}
  \Delta m_r^2(k)\geq p_{0,\text{max}}^2-\vec p^2 +k^2 x_R-
  \frac{k^4}{4}\frac{x_I^2}{p_{0,\text{max}}^2}\,,
\end{equation}
which is obtained by equating real and imaginary parts of
\eq{eq:parabola} to $x_R$ or $x_I$ respectively.  This condition keeps
the given pole outside the integration domain by shifting the parabola
to the right. Putting together the constraints from the different
poles it remains to parameterise an appropriate function which
satisfies all of them, using the simple Ansatz \eq{eq:ansatzdeltag}
for a scale-dependent $\Delta m_r^2(k)$ or alternatively using a
constant $\Delta m_r^2$ as discussed in
App.~\ref{app:constantdeltag}. This procedure is illustrated
exemplarily for the double exponential regulator, $m=2$, and
$p_{0,\text{max}}=0.33 \Lambda_{\text{UV}}$ in
\Fig{fig:deltagconstraints}. The parameter $M^2$ serves as an external
parameter in this analysis and one has to select the value for $M^2$
which constrains $\Delta m_r^2$ most strongly.

Summarising the discussion of this appendix, it is always possible to
prevent regulator poles from entering the domain
$\mathcal{S}_{p_{0,\text{max}}}$ at a given scale $k$ by introducing a
sufficiently large term $\Delta m_r^2(k)$.
\subsection{Application to the $O(N)$ model: Flow equations and numerical procedure}
\label{app:deltagflows}
\begin{figure}[t]
\centering
\includegraphics[width=\columnwidth]{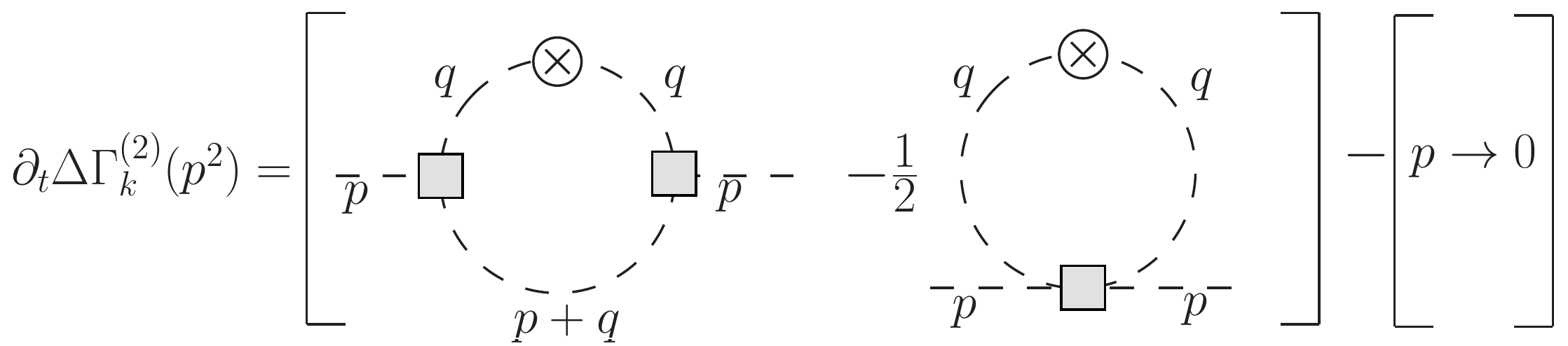}
\caption{Flow equation for the momentum-dependent part of the inverse propagator.}
\label{fig:deltaGamma2Next}
\end{figure}

In this appendix we discuss the flow equations and the numerical
solution procedure that lead to the numerical results on the mesonic
spectral functions presented in Sec.~\ref{sec:ON}.  In the truncation
presented here the momentum- and complex-frequency-dependent two-point
functions are solved for on the basis of a given solution for the
effective potential that serves as input for the three-point vertices
occurring in the flow equation for the two-point function. In the LPA
the flow equation for the effective potential takes the form
\begin{equation}
\partial_t V_k(\phi^2)=\frac{1}{2} I^{(1)}(m_\sigma^2)+\frac{N-1}{2} I^{(1)}(m_\pi^2)\,,
\end{equation}
where $m_\pi^2=2 V'_k$, $m_\sigma^2=2 V'_k+4\phi^2 V''_k$ and
\begin{equation}
  I^{(n)}(m^2)=\sumint_q \frac{\partial_t R_{k;\Delta m_r^2}(q^2)}{(
    q^2+m^2+R_{k;\Delta m_r^2}(q^2))^n}\,,
\end{equation}
which was evaluated numerically for the regulator
\eq{eq:reggeneralpurpose}. The flow equation for the Euclidean two-point
functions, see \Fig{fig:deltaGamma2Next}, is given by
\cite{Kamikado:2013sia}
\begin{align}
\label{eq:euclideangamma2flows}
\partial_t \Gamma^{(2)}_{\pi}(p)=& \left(\Gamma^{(3)}_{\sigma \pi \pi}\right)^2 (J_{\sigma\pi}(p)+J_{\pi\sigma}(p))\nonumber\\
&+\text{tadpole-terms}\,,\nonumber\\[2ex]
\partial_t \Gamma^{(2)}_{\sigma}(p)=&(N-1)\, \left(\Gamma^{(3)}_{\sigma \pi \pi}\right)^2 J_{\pi\pi}(p)+\left(\Gamma^{(3)}_{\sigma \sigma \sigma}\right)^2 J_{\sigma\sigma}(p)\nonumber\\
&+\text{tadpole-terms}\,,
\end{align}
with $\Gamma^{(3)}_{\sigma \pi \pi}$ and $\Gamma^{(3)}_{\sigma \sigma \sigma}$ given in \eq{eq:verticesfrompot} and
\begin{align}
J_{ij}(p)=&\sumint_q \partial_t R_{k;\Delta m_r^2}(q^2)\left(q^2+m_j^2+R_{k;\Delta m_r^2}(q^2)\right)^{-2}\nonumber\\
&\times\left((q+p)^2+m_i^2+R_{k;\Delta m_r^2}((q+p)^2)\right)^{-1}\,.
\end{align}
The frequency routing is chosen such that the cutoff line only
depends on the loop frequency $q_0$, that is $G\dot R G(q)$. As in
\cite{Helmboldt:2014iya} we only resolve the genuine momentum
dependence of the propagator $\Delta \Gamma^{(2)}$, i.e.\ we solve
flow equations $\partial_t \Delta \Gamma^{(2)}(p)\equiv\partial_t
\Gamma^{(2)}(p)-\partial_t \Gamma^{(2)}(0)$, for which the tadpole
terms, which are momentum-independent in our truncation, cancel.  The
full propagator is then obtained via
\begin{equation}
  \Gamma_i^{(2)}(p)=\Delta \Gamma^{(2)}_i(p)+
  \frac{\partial^2}{(\partial \phi_i)^2} V(\phi^2)\,.
\end{equation}
We use \eq{eq:EuclideanRetardedGamma2} in \eq{eq:euclideangamma2flows}
to set up flow equations for the real and imaginary parts of the
retarded propagator, keeping a fixed finite parameter $\epsilon$. Note
that in addition to this contribution it remains to track the
positions of the physical poles and to add the corresponding
contributions involving residues at their positions, see the extensive
discussion in Sec.~\ref{sec:poleprocedures}.

\subsection{Effective cutoff scales}
\label{app:effcutoffscales}
The inclusion of $\Delta m_r^2(k)$ in the regulator's shape function,
as discussed in App.~\ref{app:scaledepdeltag}, leads to a modification
of the effective cutoff scale with important physical
consequences. One way of defining such an effective cutoff scale is to
consider the gap in the propagator, i.e.\
\begin{equation}
\label{eq:keffmin}
  k_\text{eff;min}^2(k)=\min_p \left(\Gamma^{(2)}_k(p^2)
+R_k(p^2)\right)\,.
\end{equation}
Note that \eq{eq:keffmin} defines a physical cutoff scale which does
not go to zero in a massive theory. In the massless limit for
regularised propagators that depend monotonously on momentum
\eq{eq:keffmin} coincides with the effective cutoff scale
$k_{\text{eff};R(0)}$ defined via
\begin{equation}
\label{eq:keff}
k_{\text{eff};R(0)}^2(k) =\lim_{p^2\to 0} R_{k;\Delta m_r^2}(p^2)=\Delta m_r^2 r(\Delta m_r^2/k^2)\,,
\end{equation}
which can be easily evaluated for a given $\Delta
m_r^2(k)$. $k_{\text{eff};R(0)}^2$ tends to $k^2$ for $\Delta
\hat{m}_r^2\equiv \Delta m_r^2/k^2\ll 1$ and it drops exponentially with
$\Delta \hat{m}_r^2$ in the regime where $\Delta \hat{m}_r^2\gg 1$. An
alternative definition is given by
$k_{\text{eff};R=\Delta\Gamma^{(2)}}$ which defines the effective
cutoff scale as the momentum scale where the regulator gets of the
order of the propagator itself, i.e.\
\begin{equation}
\label{eq:keff2}
R_{k;\Delta m_r^2(k)}(p^2)=\Delta\Gamma^{(2)}_k(p^2)
\Biggr|_{p^2=k^2_{\text{eff};R=\Delta\Gamma^{(2)}}(k)}\,,
\end{equation}
where
$\Delta\Gamma^{(2)}_k(p^2)=\Gamma^{(2)}_k(p^2)-\Gamma^{(2)}_k(0)$. For
$\Delta m_r^2=0$ and the double exponential regulator, $m=2$,
$k_{\text{eff};R=\Delta\Gamma^{(2)}}\approx 0.86\, k$. As visible from
\Fig{fig:effcutoff} both definitions show the exponential drop of the
effective cutoff scale in the regime where $\Delta \hat{m}_r^2\gg
1$. 

The relation $k(k_\text{eff})$ can now be inverted at least
numerically in order to yield a relation $k(k_\text{eff})$, which in
turn could be used to rewrite the flow equation in terms of
$k_\text{eff}$ for numerical convenience. Effective cutoff scales as
defined in this appendix can be used to adjust relative cutoff scales
for fermions and bosons and theories with bosonic and fermionic
species.

\begin{figure}[t]
\centering
\includegraphics[width=\columnwidth]{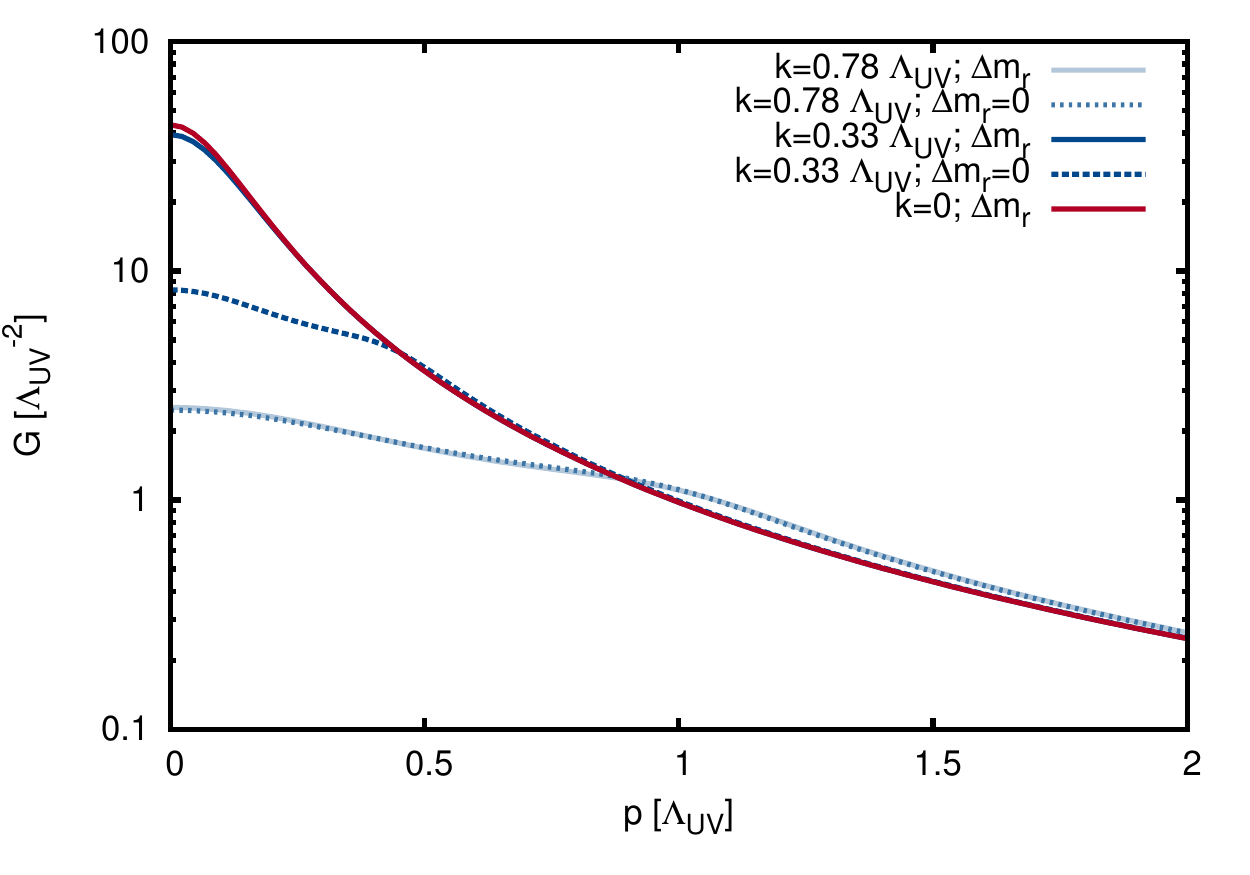}
\caption{Regularised pion propagator at different cutoff scales in a
calculation with (solid) and without (dashed) $\Delta m_r^2$-term using the same parameter
set as in \Fig{fig:deltagconstraints} in comparison to the result at $k=0$.}
\label{fig:props}
\end{figure}

We illustrate the different effective cutoff scales at the example
of the scale dependence of the regularised pion propagator shown 
in \Fig{fig:props} comparing the results of a calculation with 
and without $\Delta m_r^2$-term. For large cutoff scales $\Delta m_r^2$
is practically zero, the effective cutoff scale coincides with
the cutoff scale $k$ and both propagators agree. At smaller cutoff
scales where $\Delta m_r^2$ takes a nonvanishing value, the
propagators start to deviate. Note in particular that the 
$\Delta m_r^2$-propagator reaches its IR value already at a finite
value of the cutoff scale as a result of the exponential drop of the
effective cutoff scale with $\Delta \hat{m}_r^2$, see \Fig{fig:effcutoff}.

\section{Flow equations in the CTP formalism}
\label{app:ctpflows}
In this appendix we discuss the derivation of the flow equations for
the effective potential and the spectral functions in the CTP
formalism in detail.
\subsection{CTP formalism}
\label{app:ctpformalism}
Before discussing the CTP flow equations we start by fixing the formalism
and conventions used in this work. The main object in the following is the
matrix-valued two-point function with components defined by
\begin{align}
G_{+-}(x,y)&=-\imag\langle\phi(x)\phi(y)\rangle\,,\nonumber\\[2ex]
G_{-+}(x,y)&=-\imag\langle\phi(y)\phi(x)\rangle\,,\nonumber\\[2ex]
G_{++}(x,y)&=-\imag\langle \CT\phi(x)\phi(y)\rangle\,,\nonumber\\[2ex]
G_{--}(x,y)&=-\imag\langle\tilde\CT \phi(y)\phi(x)\rangle\,,
\end{align}
where $\CT$($\tilde \CT$) denotes (anti-)time ordering. In thermal
equilibrium and in momentum space the propagators are given solely
in terms of the spectral function, c.f.\ \cite{Das:1997gg} for the
free case,
\begin{align}
  G_{\pm\pm}(p) =& \pm\mathcal{P}\int_{-\infty}^{\infty}
  \frac{\d \omega}{2\pi}\frac{\rho(\omega,\vec p)}{p_0-\omega}\nonumber\\
  &-\imag \left(n(p_0)+\frac{1}{2}\right)\rho(p_0,\vec p)\,,\nonumber\\[2ex]
  G_{+-}(p)=&-\imag\, n(p_0)\rho(p)\,,\nonumber\\[2ex]
  G_{-+}(p)=&-\imag\, (n(p_0)+1)\rho(p)\,,
\label{eq:G_++/--}
\end{align}
where $\mathcal{P}$ denotes the Cauchy principal value and $n$ the
bosonic thermal distribution function.

In order to calculate spectral functions we require flow equations for
the real and imaginary parts of the inverse diagonal propagator
$\bar\Gamma^{(2)}(p)$ \cite{Kobes:1985kc}.  These are related to
$\Gamma^{(2)}_{++}(p)$ and $\Gamma^{(2)}_{+-}(p)$ via relations
obtained in analogy to \cite{LeBellac:2000} using \cite{Xu:1995np} to
diagonalise the CTP propagator and self-energy
\begin{align}
  \nonumber \text{Re}\, \Gamma^{(2)}_{++}(p)&=\text{Re}\,
  \bar\Gamma^{(2)}(p)\,,\\[2ex]
\nonumber \text{Im}\, \Gamma^{(2)}_{++}(p)&=\sign(p^0)\,(1+2n(p^0))\text{Im}\,\bar\Gamma^{(2)}(p)\,,\\[2ex]
  \Gamma^{(2)}_{+-}(p)&=-\imag\,\, \sign(p^0)\, 2
  n(p_0)\text{Im}\,\bar\Gamma^{(2)}(p)\,.
\label{eq:Gammabar}
\end{align}
The real and imaginary parts of the diagonalised real time two-point
function are related to the retarded correlation function via
\cite{Xu:1995np,LeBellac:2000}
\begin{align}
\label{eq:relgammabargammar}
\text{Re}\,\Gamma^{(2)}_\text{R}(p)&=\text{Re}\,\bar\Gamma^{(2)}(p)\,,\nonumber\\[2ex]
\text{Im}\,\Gamma^{(2)}_\text{R}(p)&=\sign(p_0)\,\text{Im}\,\bar\Gamma^{(2)}(p)\,.
\end{align}
Now consider the retarded propagator that, following \eq{eq:G_++/--},
is given by
\begin{align}
  G_{\text{R}}(p)&=G_{++}(p)-G_{+-}(p)\nonumber\\[2ex]
  &=\int_{-\infty}^\infty\frac{\d\omega}{2\pi}
  \frac{\rho(
    \omega,\vec p)}{p^0-\omega+\imag\epsilon}\nonumber\\[2ex]
  &=\mathcal{P}\int_{-\infty}^\infty\frac{\d\omega}{2\pi}
  \frac{\rho(\omega,\vec p)}{p^0-\omega}-\frac{\imag}{2}\rho(p^0,\vec
  p)\,.
\end{align}
This immediately implies that
\begin{align}
\label{eq:spectraldefapp}
\rho(p)&=-2\,\text{Im}\, G_{\text{R}}(p)=-2\,\text{Im}\,[\Gamma_\text{R}^{(2)}]^{-1}\nonumber\\[2ex]
&=\frac{2\,
  \text{Im}\,\Gamma^{(2)}_\text{R}(p)}{(\text{Im}\,\Gamma^{(2)}_\text{R}(p))^2+(\text{Re}\,
  \Gamma^{(2)}_\text{R}(p))^2}\,,
\end{align}
and leaves us with the determination of flow equations for
$\text{Re}\, \Gamma^{(2)}_{++}$ and for example $\Gamma^{(2)}_{+-}$ in
order to obtain the spectral function.
\subsection{CTP flow equation}
\label{app:ctpfloweq}
In this appendix we fix the conventions that are required to derive the
CTP flow equation. These are chosen such that the functional relations
between the different generating functionals stay as close to the
Euclidean relations as possible. We start by defining a generating
functional
\begin{align}
  Z_k[J_+,J_-]=&\int\mathcal{D}\varphi_+\mathcal{D}\varphi_- \exp\Biggl[\imag 
S[\varphi_+,\varphi_-]\nonumber\\
  &+\imag \Delta S_k[\varphi_+,
\varphi_-]-\imag \int_x \varphi_a(x)\eta^{ab} J_b(x)\Biggr]\,,\nonumber\\[2ex]
  \Delta S_k[\varphi_+,\varphi_-]=&\frac{1}{2}\int_{x,y} \varphi_a(x)
  \eta^{ab} \left(R_k\right)_{bc}(x,y) \eta^{cd}\varphi_d(y)\,,
\end{align}
where $\eta=\text{diag}(1,-1)$ and $a,b,c,d\in\{+,-\}$. Defining the generating functional
for connected Greens functions $W_k$ and the effective action $\Gamma_k$ via
\begin{align}
W_k[J_+,J_-]&=\imag \log Z_k[J_+,J_-]\,,\nonumber\\[2ex]
\Gamma_k[\phi_+,\phi_-]&=J_a \eta^{ab}\phi_b- W_k-\Delta S_k[\phi_+,\phi_-]\,,
\end{align}
leads to the relations
\begin{align}
\frac{\delta W_k}{\delta J_a}&=\eta^{ab}\phi_b\,,\nonumber\\[2ex]
\frac{\delta(\Gamma_k+\Delta S_k)}{\delta \phi_a}&=\eta^{ab}J_b\,,
\end{align}
and finally to the flow equation
\begin{align}
  \partial_t \Gamma_k&=\frac{\imag}{2} \text{Tr}\, \left(
\0{1}{\Gamma_k^{(2)}+R_k}\right)_{ab}(\partial_t R_k)_{ba}\nonumber\\[2ex]
  &=\frac{\imag}{2}\text{Tr}\,G_{ab}\, (\partial_t R_k)_{ba}\,.
\end{align}
\subsection{Flow equation for the effective potential}
\label{app:ctpeffpot}
As a first step we consider the flow equation for the effective
potential in the CTP formalism. Note that evaluating $\Gamma$ for
constant fields $\phi^+=\phi^-=\phi$ yields a contribution with a
vanishing real part. In order to project onto the effective potential
we take the real part after taking a single $\phi^+$-derivative and
setting fields $\phi^+=\phi^-=\phi$ to constant afterwards.  We find
\begin{align}
  \partial_{\phi^+}\partial_t \Gamma_k=\tilde \partial_t
  \0{\imag}{2}\Tr\,G_{ab}\Gamma_{ba+}\,, 
\label{eq:+flowapp}\end{align}
with the partial $t$-derivative at fixed $\Gamma_k^{(n)}$, 
\begin{align}\label{eq:dtdilde}
  \tilde\partial_t=\partial_t|_{\Gamma_k^{(n)}} = \Tr \dot R_k \0{\delta}{\delta R_k} = 
\sum_{a=\pm}\Tr \dot R^a_k \0{\delta}{\delta R^a_k}\,. 
\end{align}
In the following we resort to the canonical choice $R_k^+=-R_k^-=R_k$ and 
restrict ourselves to the effective potential. Now we have 
\begin{align}
  \partial_t \partial_\phi V_k(\phi^2)&= -\012 \dot R_k \partial_{R_k} \Tr
  \left(\text{Im}\,G_{ab}\Gamma^{(3)}_{ba+}\right)\,,
\end{align}
with $a,b=\pm$. The local potential approximation (LPA) as simplest
approximation involves inserting free spectral functions
\begin{align}
\label{eq:freeSF}
\rho(p_0,\vec p)&=(2\pi)\, \text{sign}(p_0)\,
\delta(-p_0^2+\vec p^2+m^2)\nonumber\\[2ex]
&=\frac{2\pi}{2\epsilon_p}\left(\delta(p_0-
  \epsilon_p)-\delta(p_0+\epsilon_p)\right)\,,
\end{align}
with $\epsilon_p$ from \eq{eq:epsilon}. In the LPA approximation we
have a deformed classical dispersion with a field-dependent mass as
given by
\begin{align}\label{eq:mass}
m^2_k(\phi^2)= 2 V_k'(\phi^2) +4 \phi^2 V_k''(\phi^2)\,,
\end{align} 
for a single field mode, where primes denote derivatives with respect
to $\phi^2$. Correspondingly for the regulator \eq{eq:p2p0}, the spectral function \eq{eq:freeSF}
generalises to
\begin{align}
  \rho(\omega,\vec p)=&(2\pi)\,\text{sign}(p_0)\nonumber\\
  &\times\delta\Bigl(-p_0^2+\vec p^2+m_k^2(\phi^2)\nonumber\\
&\quad+R_k(-p_0^2+\vec p^2)(r_0(p_0^2/k^2)+r_s(\vec p^2/k^2))\Bigr)\,.
\end{align}
Here, we restrict ourselves to spatial regulators. 
In the simplest case of a spatial regulator function, i.e.\ for
$R_k(p^2)\equiv k^2$, $r_0\equiv 0$ and $r_s(x)=x r(x)$ in
\eq{eq:p2p0}, we can trivially perform the frequency integration and
the flow equation for the effective potential now takes the simple
form
\begin{align}\nonumber 
  \partial_t \partial_\phi V_k(\phi)&=\012 \int_{\vec
    q}\dot{r}\partial_{r}\frac{1+2n(\epsilon_q)}{2\epsilon_q}
  \Gamma^{(3)}_{\phi\phi\phi}\\[2ex]\nonumber &=\012 \int_{\vec q}
  \vec q^2 \dot{r} \partial_{\epsilon_q}\left(\frac{1
      +2n(\epsilon_q)}{2\epsilon_q}\right)\frac{\partial
    \epsilon_q}{\partial\phi}\\[2ex]
  &=\partial_\phi \012 \int_{\vec q}\frac{\vec q^2 \dot{r}}{2\epsilon_q}\left(1+2 n(\epsilon_q)\right)\,,
\label{eq:flow3d}
\end{align}
with
\begin{align}
  \epsilon_q(\phi^2)= \sqrt{\vec q^2(1+r(\vec q^2/k^2))+m_k^2(\phi^2)}\,.
\label{eq:epsilon_k}\end{align} 
The flow equation \eq{eq:flow3d} is nothing but the derivative of the
standard flow equation for the effective action/potential for a
three-dimensional spatial regulator. In particular, for the flat
regulator, the 3d-analogue of the LPA-optimised regulator \cite{Litim:2001up}, where $r(x)=(-1+1/x)\theta(1-x)$, we arrive at
\begin{align}
  \partial_t \partial_\phi V_k(\phi^2) =&\partial_\phi \left( \0{k^5}{6
      \pi^2}\frac{\coth\left(\frac{\sqrt{k^2+m_k^2(\phi^2)}}{2T}\right)}{
      \sqrt{k^2+m_k^2(\phi^2)}}\right)\,,
\label{eq:Vk}\end{align}
the well-known flow for the spatial flat regulator as derived in the
imaginary time formalism.

\subsection{Flow equations for spectral functions}
\label{app:ctpspectral}
In the CTP formalism with spatial regulator functions the direct
computation of real time momentum and frequency dependence is
straightforward. Note however, that the regulator explicitly breaks
Lorentz invariance and frequency integrations only have polynomially
decaying integrands.

For $R_k^+=-R_k^-=R_k$ again, we find the symbolic flow equations for
the different components of the two-point function
\begin{align}
  \partial_t \Gamma^{(2)}_{++}(p)=&-\frac{\imag}{2}\tilde \partial_t
  \Tr\left(\Gamma_{+}^{(3)}G_{++}\Gamma^{(3)}_{+} G_{++}\right)\nonumber\\
  &+\frac{\imag}{2}\tilde \partial_t \Tr\left(\Gamma^{(4)}_{+} G_{++}
  \right)\,,\label{eq:gammapp}\\[2ex]
 \partial_t \Gamma^{(2)}_{+-}(p)=&-\frac{\imag}{2}\tilde \partial_t
 \Tr\left(\Gamma^{(3)}_{+}G_{+-}\Gamma^{(3)}_{-} G_{-+}
 \right)\,.\label{eq:gammapm}
\end{align}
Evaluating \eq{eq:gammapm} and neglecting the tadpole terms leads to
\begin{align}
  &\partial_t\, \text{Re}\,\Gamma^{(2)}_{++}(p) \nonumber \\[2ex]
  =&+\012 \tilde \partial_t \int_q \Gamma^{(3)}_{+}\text{Im}\,
  G_{++}(q+p)
  \Gamma^{(3)}_{+}\,\text{Re}\,G_{++}(q)\nonumber\\
  &+\012 \tilde \partial_t\int_q \Gamma^{(3)}_{+}\text{Re}\,
  G_{++}(q+p)\,
  \Gamma^{(3)}_{+}\text{Im}\,G_{++}(q)\nonumber\\[2ex]
  =&-\012 \tilde \partial_t\int_{\vec q}\frac{\Gamma^{(3)}_{+}
    \Gamma^{(3)}_{+}}{4\epsilon_q \epsilon_{q+p}}\Biggl[\left(n(
    \epsilon_{q+p})-n(\epsilon_{q})\right)\nonumber\\
  &\quad\times\mathcal{P}\left(\frac{1}{p_0-\epsilon_q+
      \epsilon_{q+p}}-\frac{1}{p_0+\epsilon_q-\epsilon_{q+p}}
  \right)\nonumber\\
  &+\left(1+n(\epsilon_{q+p})+n(\epsilon_{q})\right)\nonumber\\
  &\quad\times\mathcal{P}\left(\frac{1}{p_0-\epsilon_q
      -\epsilon_{q+p}}-\frac{1}{p_0+\epsilon_q+\epsilon_{q+p}}\right)\Biggr]\,.
\end{align}
Evaluating \eq{eq:gammapm} using the identity
$(1+n(q^0))n(q^0+p^0)=n(p^0)(n(q^0)-n(q^0+p^0))$ we find
\begin{align}
  &\partial_t \Gamma^{(2)}_{+-}(p)\nonumber\\[2ex]
  =&-\frac{\imag}{2}\tilde \partial_t\!\int_q \Gamma^{(3)}_{+}
  G_{+-}(q+p)\Gamma^{(3)}_{-} G_{-+}(q)\nonumber\\[2ex]
  =&\frac{\imag}{2} n(p^0) \tilde\partial_t\int_q
  \Gamma^{(3)}_{+}\Gamma^{(3)}_{-}(n(q^0)-n(p^0+q^0))
  \rho(q+p)\rho(q)\label{eq:preflow+-}
\end{align}
leading to  
\begin{align}
\partial_t \Gamma^{(2)}_{+-}(p)  =&\frac{\imag}{2} n(p^0) \tilde\partial_t \int_{\vec q}
  \frac{2\pi}{4 \epsilon_{q+p} \epsilon_q}\Gamma^{(3)}_{+}
  \Gamma^{(3)}_{-}\nonumber\\
  &\times\Bigl[\left(n(\epsilon_{q})-n(\epsilon_{q+p})
  \right)\nonumber\\
  &\times\left(\delta(p_0+\epsilon_q-\epsilon_{q+p})-
    \delta(p_0-\epsilon_q+\epsilon_{q+p})\right)\nonumber\\
  &+\left(1+n(\epsilon_{q})+n(\epsilon_{q+p})\right)\nonumber\\
  &\times\left(\delta(p_0-\epsilon_q-\epsilon_{q+p})-
    \delta(p_0+\epsilon_q+\epsilon_{q+p})\right)\Bigr]\,.
\label{eq:flow+-}\end{align}
If one now exploits the fact that the regulator enters the flow only
via $\epsilon_s$ one may rewrite $\tilde \partial_t$ as
\begin{equation}
\int_{\vec s} \dot{R}(\vec s)\0{\delta}{\delta R(\vec s)}=\int_{\vec s}
  \0{\dot{R}(\vec s)}{2\epsilon_s}\0{\delta}{\delta \epsilon_s}\,.
\end{equation}
For general regulators one now can perform the $\epsilon$-derivative
and then resolve the $\delta, \delta'$-functions. Here we restrict
ourselves again to the three-dimensional flat regulator
\begin{equation}\label{eq:3dflat}
R_{k}^{\text{\tiny{flat}}}(\vec p^2)= (k^2 -\vec p^2)\theta(k^2 -\vec p^2)\,. 
\end{equation} 
This allows us to analytically resolve the $\delta$-functions before
taking the $\tilde t$-derivative with $\tilde \partial_{t}
\Gamma_k^{(n)}=0$ for all $n\in\N^+$, i.e.\ which acts only on the
explicit $k$ dependence of the propagators.  The $\delta$-functions in
\eq{eq:flow+-} read
\begin{align}
  & \delta(p_0\pm\epsilon_q \pm \epsilon_{q+p})  \nonumber\\[2ex]
  =& \delta(p_0\pm \epsilon_{k} \pm \epsilon_k) \theta(k^2-\vec q^2)
  \theta(k^2 - (\vec q+\vec p)^2) \nonumber \\[1.5ex]
  &+ \delta(p_0\pm \epsilon_{q} \pm \epsilon_{k} ) \theta(\vec q^2-k^2)
  \theta(k^2 - (\vec q+\vec p)^2) \nonumber \\[1.5ex]
  &+ \delta(p_0\pm \epsilon_{k} \pm \epsilon_{q+p} ) \theta(k^2-\vec
  q^2)
  \theta(  (\vec q+\vec p)^2-k^2)\nonumber \\[1.5ex]
  &+ \delta(p_0\pm \epsilon_{q} \pm \epsilon_{q+p} ) \theta(\vec
  q^2-k^2) \theta( (\vec q+\vec p)^2-k^2)\,,
  \label{eq:rewritedelta} 
\end{align}
where both signs can be varied independently.
The $\delta$-function in the first term on the right hand side does
not depend on the loop momentum $\vec q$, however, the integration can
readily be performed. The other three $\delta$-functions in
\eq{eq:rewritedelta} depend on the loop momentum $q$.  They can be
rewritten in terms of $\delta$-functions w.r.t.\ $q$ with the help of
\begin{align}
   \delta(p_0+\epsilon_q \pm \epsilon_{k}) = & \0{\epsilon_{q}
     \delta(q-q^{(1)\pm}) }{q}  \,,\nonumber \\[2ex]
   \delta(p_0+\epsilon_{q+p} \pm \epsilon_{k}) = & \sum_{i=1}^2
   \0{\epsilon_{q+p}\delta(q-q^{(2)\pm}_i)}{| q+p\,
     x|}    \,,\nonumber \\[2ex]
   \delta(p_0+\epsilon_q\pm \epsilon_{q+p})= & \sum_{i=1}^2 \0{
     \delta( q-q^{(3)}_{i} )}{| \0{q}{\epsilon_q} \pm \0{(q+p\,
       x)}{\epsilon_{q+p}}|}\,,
  \label{eq:resolvedelta} 
\end{align}
where $x=\vec q\cdot \vec p/(|\vec q|| \vec p|)$ and we can assume without
loss of generality that the second summand appears with positive coefficient. 
Note that $q^{(i)}$ depend on $p_0, |\vec p|, x, m_k^2,k$.  The $\tilde t$-derivative only
hits the explicit $k$ dependence and not $m_k^2$. The $q^{(i)}$ are
given analytically, to wit
\begin{align}
  q^{(1)\pm}=& \sqrt{(p_0\pm\epsilon_k)^2-m^2}\,, \nonumber \\[1.5ex]
  q^{(2)\pm}_i= & -|\vec p| x+(-1)^i\sqrt{\vec p^2 (x^2-1)+(p_0\pm
    \epsilon_k)^2-m^2}\,,\nonumber \\[1.5ex]
  q_{i}^{(3)\phantom{\pm}} = & \0{1}{2\left(-p_0^2+\vec p^2 x^2\right)}\Bigl\{
  -|\vec p|^3 x+|\vec p| p_0^2 x \nonumber\\[1.5ex]
  & +(-1)^i p_0 \sqrt{(\vec p^2-p_0^2)^2+4m^2(-p_0^2+ \vec p^2
    x^2)}\Bigr\}\,,
\label{eq:roots} 
\end{align} 
where the set of the two roots of $q^{(3)}$ agree.

\subsection{Application to the $O(N)$ model}
\label{app:ctponmodel}
As nontrivial example we discuss the computation of spectral functions
in the $O(N)$ model on the basis of a given solution for the effective
potential.  As demonstrated above for a single scalar field the real
time flow equation for the effective potential coincides with that
derived in the imaginary time formalism. Turning to two-point
functions, the application to the $O(N)$ model requires more general
expressions with different particles in the loop diagrams. Therefore
we consider
\begin{align}\bar
J_{ji}^{\text{Re}}(p)=&+\frac{1}{2}\int_q\text{Im}\,G^j_{++}(q+p)\,
\text{Re}\,G^i_{++}(q)\nonumber\\
&+\frac{1}{2}\int_q\text{Re}\,G^j_{++}(q+p)\,\text{Im}\,
G^i_{++}(q)\,,
\label{eq:preJijre}\end{align} 
leading to 
\begin{align}
\bar
J_{ji}^{\text{Re}}(p)=&-\frac{1}{2}\int_{\vec{q}}\frac{1}{4\epsilon^i_q
  \epsilon^j_{q+p}}\Biggl[\left(n(\epsilon^i_{q})-n(
  \epsilon^j_{q+p})\right)\nonumber \\
&\quad\times\mathcal{P}\left(\frac{1}{p_0+
    \epsilon^i_q-\epsilon^j_{q+p}}-\frac{1}{p_0-\epsilon^i_q
    +\epsilon^j_{q+p}}\right)\nonumber\\
&+\left(1+n(\epsilon^i_{q})+n(\epsilon^j_{q+p})\right)
\nonumber\\
&\quad\times\mathcal{P}\left(\frac{1}{p_0-
    \epsilon^i_q-\epsilon^j_{q+p}}-\frac{1}{p_0+
    \epsilon^i_q+\epsilon^j_{q+p}}\right)\Biggr]\,.
\label{eq:Jijre}\end{align}
and similarly for the imaginary part, following
\eq{eq:Gammabar},
\begin{align}
  \bar J^{\text{Im}}_{ji}(p)=&-\frac{1}{2}\frac{\sign(p^0)}{2 n(p_0)}\int_qG^j_{+-}(q+p)G^i_{-+}(q)\nonumber\\[2ex]
  =&\frac{1}{4}\sign(p_0)\nonumber\\
  &\times\int_q\left(n(q^0)-n(q^0+p^0)\right)\rho^j(q+p)\rho^i(q)\,,
\label{eq:preJijim}\end{align}
leading to 
\begin{align}
  \bar J^{\text{Im}}_{ji}(p)=&
\pi\,\sign(p_0)\int_{\vec q}\frac{1}{8 \epsilon^j_{q+p} \epsilon^i_q}\nonumber\\
  &\Bigl[\left(n(\epsilon^i_{q})-n(\epsilon^j_{q+p})\right)\nonumber\\
  &\times\left(\delta(p_0+\epsilon^i_q-\epsilon^j_{q+p})-\delta(p_0-\epsilon^i_q+\epsilon^j_{q+p})\right)\nonumber\\
  &+\left(1+n(\epsilon^i_{q})+n(\epsilon^j_{q+p})\right)\nonumber\\
  &\times\left(\delta(p_0-\epsilon^i_q-\epsilon^j_{q+p})-\delta(p_0+\epsilon^i_q+\epsilon^j_{q+p})\right)\Bigr]\,.
\label{eq:Jijim}\end{align}
where the subscripts $i(j)$ refer to masses $m_i^2(m_j^2)$ in the
respective propagators $G^i$, spectral functions $\rho^i$ and
quasiparticle energies $\epsilon^i$. Using these definitions we can
write the flow equations for the real and imaginary parts of the
respective inverse propagators as
\begin{align}
\label{eq:floweqspectralrealtimeon}
\partial_t \text{Im}\,\bar\Gamma^{(2)}_{\pi\pi}(p)=& \left(\Gamma^{(3)}_{\sigma \pi \pi}\right)^2 \tilde\partial_t (\bar J^{\text{Im}}_{\sigma\pi}(p)+\bar J^{\text{Im}}_{\pi\sigma}(p))\nonumber\\[2ex]
\partial_t \text{Re}\,\bar\Gamma^{(2)}_{\pi\pi}(p)=&\left(\Gamma^{(3)}_{\sigma \pi \pi}\right)^2 \tilde\partial_t (\bar J^{\text{Re}}_{\sigma\pi}(p)+\bar J^{\text{Re}}_{\pi\sigma}(p))+\text{const.}\nonumber\\[2ex]
\partial_t \text{Im}\,\bar\Gamma^{(2)}_{\sigma\sigma}(p)=&(N-1)\, \left(\Gamma^{(3)}_{\sigma \pi \pi}\right)^2 \tilde\partial_t \bar J^{\text{Im}}_{\pi\pi}(p)\nonumber\\
&+\left(\Gamma^{(3)}_{\sigma \sigma \sigma}\right)^2 \tilde\partial_t \bar J^{\text{Im}}_{\sigma\sigma}(p)\nonumber\\[2ex]
\partial_t \text{Re}\,\bar\Gamma^{(2)}_{\sigma\sigma}(p)=&(N-1)\, \left(\Gamma^{(3)}_{\sigma \pi \pi}\right)^2 \tilde\partial_t \bar J^{\text{Re}}_{\pi\pi}(p)\nonumber\\
&+\left(\Gamma^{(3)}_{\sigma \sigma \sigma}\right)^2 \tilde\partial_t \bar J^{\text{Re}}_{\sigma\sigma}(p)+\text{const.}\,,
\end{align}
identifying $\sigma\equiv 0$ and $\pi\in\mathbbm{N}_+$ in the
subscripts. As above we only resolve the genuine momentum dependence
of the real part of the propagator by considering the flow equations
$\partial_t\Delta
\Gamma^{(2)}(p)=\partial_t\Gamma^{(2)}(p)-\partial_t\Gamma^{(2)}(0)$,
for which the momentum-independent tadpole terms cancel, to which the
corresponding contribution from the effective potential is added in
order to obtain the full propagator. Here we aim to evaluate the
spectral functions on the basis of a given solution for the effective
potential. Therefore we approximate three-point functions
momentum-independently from the corresponding vertices extracted from
the effective potential, i.e.\ explicitly via
\begin{align}
  \Gamma^{(3)}_{\sigma\pi\pi}&\equiv\Gamma^{(3)}_{+;\sigma\pi\pi}=-
  \Gamma^{(3)}_{-;\sigma\pi\pi}=4 V_k'' \phi\nonumber\\[2ex]
  \Gamma^{(3)}_{\sigma\pi\pi}&\equiv\Gamma^{(3)}_{+;\sigma\sigma\sigma}=-
  \Gamma^{(3)}_{-;\sigma\sigma\sigma}=12 V_k'' \phi+8 V_k^{(3)}
  \phi^3\,,
\label{eq:verticesfrompot}
\end{align} 
where derivatives denote derivatives with respect to $\phi^2$.  As
simplest nontrivial application we are interested in zero temperature
spectral functions at vanishing external three-momentum $\vec p=\vec
0$. In this case we find
\begin{align}
\label{eq:Jreimrealtime}
\bar J^{\text{Re}}_{ji}(p_0)=&-\int_{\vec q}\frac{1}{4\epsilon^j_{q}
  \epsilon^i_q}\mathcal{P}\left(\frac{\epsilon_q^i+\epsilon^j_q}{p_0^2-
    (\epsilon_q^i+\epsilon^j_q)^2}\right)\,,\nonumber\\[2ex]
\bar J^{\text{Im}}_{ji}(p_0)=&-\pi\,\sign(p_0)\sum_\pm\int_{\vec
  q}\frac{\mp\delta(\epsilon^i_q+\epsilon^j_{q}\mp
  p_0)}{8\epsilon^j_{q} \epsilon^i_q}\,.
\end{align}
For $|\vec q|<k$ we have
$\delta(\epsilon^i_q+\epsilon^j_{q}-p_0)=\delta(\epsilon^i_k+\epsilon^j_{k}-p_0)$,
where $\epsilon^i_k=\sqrt{k^2+m_i^2}$, whereas for $|\vec q|>k$ we
find
\begin{equation}
  \delta(\epsilon^i_q+\epsilon^j_{q}-p_0)=\frac{\delta(q-q_f(p_0,m_i,m_j))}{|
    \frac{q}{\epsilon_q^i}+\frac{q}{\epsilon_q^j}|}\,,
\end{equation}
where 
\begin{equation}
q_f(p_0,m_i,m_j)=\sqrt{\frac{((m_i\!-\!m_j)^2-p_0^2)((m_i\!+\!m_j)^2-p_0^2)}{4 p_0^2}}\,.
\end{equation} 
It remains to evaluate the $\tilde t$-derivative, which acts
only on the explicit $k$ dependence. Here only $|\vec q|<k$
contributes and we find, restricting ourselves without loss of
generality to $p_0>0$,
\begin{align}
  &\tilde\partial_t \bar J^{\text{Re}}_{ji}(p_0)\nonumber\\[2ex]
  &=-\frac{k^3}{24\pi^2}\tilde\partial_t\left(\frac{1}{\epsilon_k^i
      \epsilon_k^j}\mathcal{P}\left(\frac{\epsilon_k^i
        +\epsilon_k^j}{p_0^2
        -(\epsilon_k^i+\epsilon_k^j)^2}\right)\right)\nonumber\\[2ex]
  &=-\frac{k^5}{24\pi^2}\mathcal{P}\frac{(\epsilon_k^i+
    \epsilon_k^j)^3(\epsilon_k^i{}^2+\epsilon_k^i\epsilon_k^j+
    \epsilon_k^j{}^2)-(\epsilon_k^i{}^3+\epsilon_k^j{}^3)p_0^2}{
    \epsilon_k^i{}^3\epsilon_k^j{}^3((\epsilon_k^i+
    \epsilon_k^j)^2-p_0^2)^2}\,,\nonumber\\[2ex]
  &\tilde\partial_t \bar J^{\text{Im}}_{ji}(p_0)\nonumber\\[2ex]
  &=\frac{k^3}{48\pi}\tilde\partial_t\left(\frac{ \delta(
      \epsilon_k^i+\epsilon_k^j-p_0)}{\epsilon_k^i
      \epsilon_k^j}\right)\nonumber\\[2ex]
  &=\frac{k^3}{48\pi}\frac{1}{\epsilon_k^i+\epsilon_k^j}
  \Biggl[\delta'(k-k_s^{ij}(p_0))-\frac{k}{\epsilon_k^i
    \epsilon_k^j}\delta(k-k_s^{ij}(p_0))\Biggr]\,.
\label{eq:partialtJ}
\end{align}
For the imaginary part we rewrote the occurring delta functions as
delta functions in $k$ via
\begin{equation}
  \delta(\epsilon_k^i+\epsilon_k^j-p_0)=\frac{\delta(k-
    k^{ij}_s(p_0))}{\frac{k}{\epsilon_k^i}+\frac{k}{\epsilon_k^j}}\,,
\end{equation}
where $k^{ij}_s(p_0)$ is the solution of the equation 
\begin{equation}
\epsilon_{k^{ij}_s(p_0)}^i+\epsilon_{k^{ij}_s(p_0)}^j=p_0\,.
\end{equation}
In particular, for $\epsilon_{k}^i\geq m^i_{k=0}$ this entails that no
imaginary part can build up below the threshold i.e.\ $p_0\geq
m_\pi+m_\sigma$ for the pion or $p_0\geq 2 m_\pi$ for the sigma meson
spectral function respectively. Correspondingly the spectral functions
have to vanish below these thresholds. The only exception occurs for
momenta where $\text{Re}\, \bar \Gamma^{(2)}(p)=\text{Im}\, \bar
\Gamma^{(2)}(p)=0$, see \eq{eq:spectraldef}. For $\vec p=0$ this
occurs at the pole mass where $p_0^2=m_\text{pole}^2$, see
\cite{Helmboldt:2014iya} for a discussion of different mass
definitions. Here the spectral function shows a delta function. As it
is obvious from \eq{eq:partialtJ}, the imaginary part can be computed
without even explicitly carrying out an integration in $k$.

\subsection{Equivalence of real and imaginary time flows}
\label{app:equivalence}
These results can be compared to spectral functions obtained from
Euclidean calculations with complex external momenta using spatial
regulator functions \cite{Kamikado:2013sia} in an otherwise identical
truncation. In fact, the flow equations put forward here turn out to
be formally equivalent to the ones put forward in
\cite{Strodthoff:2011tz,Kamikado:2012bt,Kamikado:2013sia,%
  Tripolt:2013jra,Tripolt:2014wra} starting from an imaginary time
formalism. This equivalence is nothing but the equivalence of using
the real time or the imaginary time formalism for the evaluation of
one-loop diagrams \cite{LeBellac:2000,Aurenche:1991hi}. It is however
worthwhile to illustrate the formal equivalence again for this
particular example, restricting ourselves for simplicity to the case
of zero temperature and vanishing spatial external momentum. We
consider the expression for the loop diagram for Euclidean external
momentum, which reads
\begin{align}
  \bar J^\text{eucl}_{ji}(p_0)=&-\frac{1}{2}\int_q G^j(q+p) G^i(q)
  \nonumber\\[2ex]
  =&-\int_{\vec q}\frac{1}{4\epsilon_q^i \epsilon_q^j}
  \frac{\epsilon_q^i+ \epsilon_q^j}{p_0^2+(\epsilon_q^i+
    \epsilon_q^j)^2}\nonumber\\[2ex]
  =& -\int_{\vec q}\frac{1}{8\epsilon_q^i
    \epsilon_q^j}\Biggl(\frac{1}{\imag
    p_0+\epsilon_q^i+\epsilon_q^j}\nonumber \\
  & \hspace{2cm}+\frac{1}{-\imag p_0+\epsilon_q^i+\epsilon_q^j}\Biggr)\,.
\end{align}
Performing the analytic continuation we find using
\eq{eq:EuclideanRetardedGamma2} and the Sokhotski-Plemelj formula
\begin{align}
\label{eq:correspondenceimagreal}
\bar J^\text{R,IT}_{ji}(p_0) =& -\lim_{\epsilon\to 0} \bar
J^\text{eucl}_{ij}(-\imag(p_0
+\imag\epsilon))\nonumber\\[2ex]
=& \mathcal{P} \int_{\vec q}\frac{1}{4\epsilon_q^i \epsilon_q^j}
\frac{\epsilon_q^i+
  \epsilon_q^j}{-p_0^2+(\epsilon_q^i+\epsilon_q^j)^2}\nonumber\\
&+\imag\pi \sum_\pm \int_{\vec q}\frac{\pm\delta(\mp p_0+
  \epsilon_q^i+ \epsilon_q^j)}{8\epsilon_q^i \epsilon_q^j}\,.
\end{align}
By comparison to \eq{eq:Jreimrealtime} one finds
\begin{align}
\text{Re}\,\bar J^\text{R,IT}_{ji}(p_0)&=\bar J_{ji}^{\text{Re}}(p_0)\nonumber\\[2ex] 
\text{Im}\,\bar J^\text{R,IT}_{ji}(p_0)&=\sign(p_0)\,\bar J_{ji}^{\text{Im}}(p_0)\,,
\end{align}
which is completely consistent with \eq{eq:relgammabargammar}. 

\bibliographystyle{bibstyle}
\bibliography{../bib_master}
\end{document}